\documentclass{aa}  

\def\heao1{{\it HEAO-1\/}}

\def\spitzer{{\it Spitzer\/}}
\def\scuba{{\it SCUBA\/}}

\def\swift{{\it Swift\/}}

\def\gsimeq{{_>\atop^{\sim}}}
\def\lsimeq{{_<\atop^{\sim}}}
\def\lesssim{\mathrel{\hbox{\rlap{\hbox{\lower4pt\hbox{$\sim$}}}\hbox{$<$}}}}
\def\gtrsim{\mathrel{\hbox{\rlap{\hbox{\lower4pt\hbox{$\sim$}}}\hbox{$>$}}}}
\newcommand{\cgs}{ ${\rm erg~cm}^{-2}~{\rm s}^{-1}$} 
\newcommand{\lum}{\rm erg~s$^{-1}$}

\def\xray{\hbox{X-ray}}
\def\submm{\hbox{sub-mm}}

\usepackage{graphicx}
\usepackage{txfonts}
\usepackage{natbib}
\usepackage{rotating}

\begin{document}
\title{The HELLAS2XMM survey}

\subtitle{XIII. Multi-component analysis \\
of the spectral energy distribution of obscured AGN}

\author{F. Pozzi,\inst{1,2}
        C. Vignali,\inst{1,2}
        A. Comastri,\inst{2}
        E. Bellocchi,\inst{3}
        J. Fritz,\inst{4}
        C. Gruppioni,\inst{2}
        M. Mignoli,\inst{2}
        R. Maiolino,\inst{5}
        L. Pozzetti,\inst{2} 
        M. Brusa, \inst{6}
        F. Fiore, \inst{5} 
        \and
        G. Zamorani\inst{2}   
}


\institute{Dipartimento di Astronomia, Universit\`a degli Studi di Bologna, 
Via Ranzani 1, I--40127 Bologna, Italy
\and
INAF --- Osservatorio Astronomico di Bologna, Via Ranzani 1, I--40127 
Bologna, Italy
\and
Instituto de Estructura de la Materia (IEM/CSIC), C/ Serrano 121, 28006 
Madrid, Spain 
\and
Sterrenkundig Observatorium, Vakgroep Fysica en Sterrenkunde, Universeit Gent,
Krijgslaan 281, S9  9000 Gent, Belgium
\and
INAF --- Osservatorio Astronomico di Roma, Via di Frascati 33, 00040
Roma, Italy
\and
Max Planck Institut f\"ur Extraterrestrische Physik (MPE), 
Giessenbachstrasse 1, D-85748 Garching bei M\"unchen, Germany  
}

\authorrunning{F. Pozzi, C. Vignali, A. Comastri et al.}
\titlerunning{SEDs of obscured AGN}



\abstract
{}
{
We combine near-to-mid-IR \spitzer\ data with shorter wavelength 
observations (optical to X-rays) to get insights on the properties 
of a sample of luminous, obscured Active Galactic Nuclei (AGN). 
We aim at modeling their broad-band Spectral Energy Distributions (SEDs) 
in order to estimate the main parameters related to the dusty torus 
which is assumed responsible for the reprocessed IR emission. 
Our final goal is to estimate the intrinsic nuclear luminosities 
and the Eddington ratios for our luminous obscured AGN.
}
{
The sample comprises 16 obscured high-redshift ($0.9{\lsimeq}z{\lsimeq}2.1$) \xray\ luminous quasars 
(L$_{2-10keV}{\sim}10^{44}$erg~s$^{-1}$) selected from the HELLAS2XMM survey in 
the 2--10~keV band. The optical-IR SEDs are described by a multi-component 
model including a stellar component to account for the optical and 
near-IR emission,  an AGN component which dominates in the mid-IR  (mainly emission from a dusty torus 
heated by nuclear radiation) and a starburst 
to reproduce the far-IR bump. A radiative transfer code to compute the 
spectrum and intensity of dust reprocessed emission was extensively tested 
against our multiwavelength data. 
While the torus parameters and the BH accretion luminosities are a direct 
output of the SED-fitting procedure, the BH masses are estimated indirectly, 
by means of the local M$_{bulge}$-M$_{BH}$ relation. 
}
{
The majority (${\sim}$80\%) of the sources show moderate optical depth
($\tau_{9.7{\mu}m}{\leq}3$) and the derived column densities N$_{H}$ are consistent with the \xray\ 
inferred values ($10^{22}
{\lsimeq}$N$_{H}{\lsimeq}3{\times}10^{23}$~cm$^{-2}$) for most of the objects, 
confirming that the sources are moderately obscured Compton-thin AGN. 
Accretion luminosities in the range $5{\times}10^{44}{\lsimeq}L_{bol} 
{\lsimeq}4{\times}10^{46}$~erg~s$^{-1}$ are inferred from the 
multiwavelength fitting procedure.
We  compare model luminosities with those obtained by integrating 
the observed SED, finding that the latter are lower by a factor of 
${\sim}$2 in the median. The discrepancy can be as high as an order of 
magnitude for  models with high optical depth ($\tau_{9.7{\mu}m}=10$). 
The ratio between the luminosities obtained by the fitting procedure 
and from the observed SED suggest that, at least for Type~2 AGN, observed 
bolometric luminosities are likely to underestimate intrinsic ones and the 
effect is more severe for highly obscured sources.
Bolometric corrections from the hard X--ray band are computed and 
have a median value of k$_{2-10{keV}}{\sim}20$. 
The obscured AGN in our sample are characterized by relatively low 
Eddington ratios (median ${\lambda}_{Edd}{\sim}0.08$).  On average, 
they are consistent with the Eddington ratio increasing at increasing  bolometric 
correction (e.g. \citealt{2009MNRAS.392.1124V}).}
{}


\maketitle

%

\section{Introduction}
\label{intro_sec}

A robust determination of Active Galactic Nuclei (AGN) Spectral Energy 
Distributions (SEDs) is of paramount importance to 
better understand the accretion processes onto supermassive black holes 
(SMBHs) and their cosmological evolution. 
According to our present knowledge, the bulk  of accretion luminosity 
is emitted in the optical-UV range with a quasi-thermal spectrum originating 
in an optically thick, geometrically thin, accretion disk. 
Electrons with temperatures of the order of a few hundreds of keV 
form a hot corona which upscatters disk photons to \xray\ energies 
with a power law spectrum and an exponential high-energy cut-off 
corresponding to the electron temperature 
(e.g., \citealt{1991ApJ...380L..51H}). 
Dusty material, possibly with a toroidal geometry, intercepts a fraction of 
the primary continuum which depends on the covering factor. 
The absorbed energy is re-emitted in the near to far infrared with a grey-body 
spectral shape. 

The SED of optically bright, unobscured QSO is relatively well known. After 
the seminal work of \cite{1994ApJS...95....1E}, fairly accurate measurements 
were published by \cite{2006ApJS..166..470R} using Sloan Digital Sky Survey 
(SDSS) data. The emission is characterized by a double bump. 
In a ${\nu}F_{\nu}$ diagram, the optical-UV spectrum rises steeply towards 
the shortest accessible wavelengths. It is commonly referred to as the 
Big Blue Bump and thought to be the accretion disk signature. The IR bump is 
weaker and likely due to a dusty torus seen almost face-on. 
The ratio between UV (at 2500~\AA) and \xray\ luminosity (at 2~keV), 
parameterized by the slope 
$\alpha_{ox}$\footnote{$\alpha_{ox}= -\frac {log (f_{2keV}/f_{2500\AA})} {log (\nu_{2keV}/\nu_{2500\AA})}$, \cite{1981ApJ...245..357Z}.} 
of the power law connecting the rest-frame luminosities, 
increases with increasing UV luminosity 
(e.g., \citealt{2003AJ....125..433V}; \citealt{2006AJ....131.2826S}).

The average SEDs for radio-loud and radio-quiet Type~1 AGN presented in 
\cite{1994ApJS...95....1E} allow the estimation of bolometric corrections, 
which are a key parameter to determining the bolometric luminosity from 
observations at a given frequency, and the Eddington ratio, once the SMBH 
mass is known. 
By including the $\alpha_{ox}$ vs. UV luminosity dependence, 
\cite{2004MNRAS.351..169M} and \cite{2006ApJS..163....1H} computed 
luminosity-dependent bolometric corrections and adopt them to estimate 
the local SMBH mass density from the observed \xray\ luminosity functions.

The luminosity dependence of bolometric corrections was recently questioned 
by \cite{2009MNRAS.392.1124V}, who pointed out the importance of 
simultaneous optical-UV and \xray\ observations and reddening corrections in 
the UV. They suggest that the bolometric correction correlates with the 
Eddington ratio rather than with the bolometric luminosity. 
Their observational results align with the predictions of 
accretion disk models (e.g., \citealt{1997MNRAS.286..848W}) where higher blue-bump to \xray\ ratios for sources with higher  
Eddington ratios are expected.

While important progress has been made towards a better determination of 
Type~1 AGN SEDs, our knowledge of Type~2 broad-band spectra is much more 
limited despite the fact that most of the accretion-driven 
energy density in the Universe is expected to occur in obscured AGN 
(e.g., \citealt{1999MNRAS.308L..39F}, \citealt{2007A&A...463...79G} 
and references therein). Therefore, a robust estimate of their bolometric 
luminosity is extremely important to properly address the issue of 
SMBH evolution over cosmic time. 
 
Nuclear accretion luminosity in Type~2 AGN is very faint in the optical-UV and 
soft X-rays. Moreover, the host galaxy stellar light often dominates 
in the optical making it difficult to disentangle nuclear emission from 
starlight.
Infrared  emission is only marginally affected by dust obscuration and 
has proved to be a powerful indicator of dust obscured AGN. 
In particular, the thermally reprocessed nuclear emission of obscured Type~2 
AGN is expected (e.g., \citealt{2006MNRAS.366..767F}, 
hereafter F06, and references within) to peak around a few tens of micron.

Mid-IR (MIR) observations, and especially those obtained in the last few years 
with the \spitzer\ satellite, are extremely efficient in the study 
of obscured AGN (e.g. \citealt{2005ApJ...627..134R}; 
\citealt{2005Natur.436..666M}; \citealt{2006ApJ...653..101W}; 
\citealt{2008ApJ...672...94F}).
In a previous paper (\citealt{2007A&A...468..603P}) we presented 
the first analysis of the mid-IR data of a \spitzer\ program devoted to 
a systematic study of the broad-band properties of \xray\ selected, 
luminous obscured quasars. 
In \cite{2007A&A...468..603P}, the SEDs were reproduced by means of 
SED templates from \cite{2004MNRAS.355..973S}.

Here we present the observational data for our final sample of 16 
obscured quasars and the detailed modeling of their broad-band SED 
using a more complete multi-component model, with goodness of fit estimated
via a $\chi^{2}$ analysis.

The outline of the paper is as follows: in $\S$2 the \xray\ selected 
quasar sample is presented, along with all the available multi-band (optical, 
near-IR (NIR) and sub-mm) and spectroscopic follow-up. 
The \spitzer\ data are presented in $\S$3, along with data reduction and 
analysis. In $\S$4, the complete multi-component model is described. 
In $\S$5, the best-fitting solutions are discussed, while in $\S$6 
we focus on the black hole physical properties that can be constrained 
from the best-fitting procedure. 
Finally, the main results are summarized in $\S$ 7. 

Hereafter, we adopt the concordance cosmology 
($H_{0}=70$~km~sec$^{-1}$~Mpc$^{-1}$, $\Omega_{m}$=0.3 and 
$\Omega_{\Lambda}$=0.7, \citealt{2003ApJS..148..175S}). 
Magnitudes are expressed in the Vega system.

\section{The sample}
\label{sample_sec}

The sample presented in this work comprises 16 \xray\ obscured quasars 
detected in the HELLAS2XMM survey (\citealt{2002ApJ...564..190B}) 
and observed by \spitzer\ in 2006.
The HELLAS2XMM survey is a shallow, large-area hard \xray\ survey 
($S_{2-10\ keV}>10^{-14}$~\cgs\ ) over a final area of 1.4~deg$^2$. 
The catalogue comprises 232 \xray\ sources; $\sim$92\% of the sample is 
optically identified down to R$\sim$25, while $\sim$70\% of the sources 
have a spectroscopic classification (\citealt{2003A&A...409...79F}, 
\citealt{2006A&A...445..457M}, \citealt{2007A&A...466...31C}).

\begin{figure}
\centering
\includegraphics[width=0.5\textwidth]{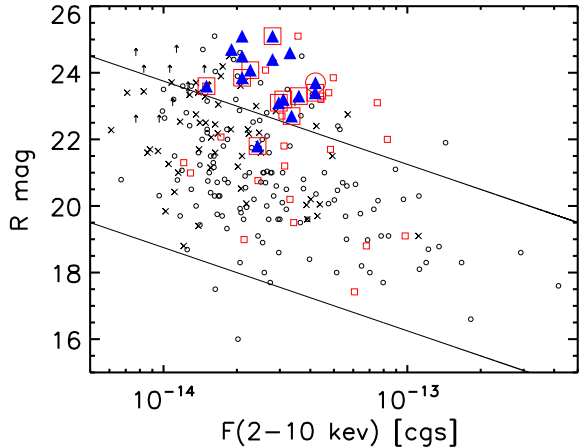} 
\caption{R-band magnitudes vs. hard \xray\ (2--10 keV) flux for the full 
HELLAS2XMM sample (\citealt{2007A&A...466...31C}). Blue triangles represent 
the sources included in the present analysis: blue triangles inside 
red symbols represent the sources with spectroscopic redshifts. 
Other symbols: empty red squares = sources spectroscopically classified as 
Type~2 AGN; empty circles =  sources spectroscopically classified as non-Type~2 
AGN (Type~1 AGN, emission-line galaxies, early-type galaxies and 
groups/clusters of galaxies); crosses: objects not observed spectroscopically; 
upward arrows = lower limits.  The dashed lines represent the 
loci of constant X/O ratio (X/O=$\pm{1}$).}
\label{fig1}
\end{figure}

The 16 sources were selected from the original survey in order to 
include the most luminous obscured quasars. The selection  was primarily based on the X-ray-to-optical flux ratio 
(hereafter X/O)\footnote{ X/O is defined as $\log{\frac{F_X}{F_R}}$. We used 
$f_{R}(0)=1.74{\times}10^{-9}$~erg~cm$^{-2}$s$^{-1}$~\AA$^{-1}$ 
and ${\Delta}{\lambda}_{R}=2200$~\AA, \cite{1990hsaa.book.....Z}.}, 
which has been proved to be an efficient way to select high-redshift 
(z${\gsimeq}$1), obscured quasars (see \citealt{2003A&A...409...79F}). 
All but one of the sources were selected to have X/O greater 
than 1 (see Fig.~\ref{fig1} and Table~\ref{table1}); the only exception is  GD~158\#19 (X/O${\sim}{0.63}$) which was included in the sample 
for its peculiar properties (see \citealt{2009MNRAS.395.2189V}, hereafter V09). We note, however, that not all of the HELLAS2XMM sources 
matching this selection criterion are present in this work. 

The selected sources are relatively faint in the optical band, with an R-band
magnitude in the range 21.8-25.1 (the brightest object being the peculiar source
GD~158\#19).  
By combining optical photometry with deep K$_{s}$-band photometry 
(obtained with the Infrared Spectrometer And Array Camera, {\it ISAAC}, 
mounted on the ESO-VLT1 Telescope), almost all the objects  are found to be 
Extremely Red Sources (EROs, R-K$_{s}{\geq}5$). The link between high X/O 
ratios and optical-to-near colours was studied by e.g., \cite{2005A&A...432...69B}. Considering different \xray\ surveys at different 
depths, they find a clear trend: the higher X/O, the redder the source. 

In \cite{2004A&A...418..827M}, detailed K$_{s}$-band morphological 
studies were presented for 8 objects of the sample selected from those sources
with the more
extreme R-K$_{s}$ colours: the majority of the sources 
(6 over 8)  have an extended K$_{s}$-band morphology, 
consistent with an elliptical-type profile, without 
evidence for a nuclear point-like source (which would be expected to trace 
the \xray\ AGN). This suggests that the nuclear emission is 
diluted and hidden by the host galaxy up to at least 2.2$\mu$m. 
In Table~\ref{table1}, the R and K$_{s}$-band magnitudes are reported.

 For 11 sources, spectroscopic information is available thanks to optical 
(8  sources, see \citealt{2003A&A...409...79F} and 
\citealt{2007A&A...466...31C}) and near-IR spectroscopy (3 sources, see 
\citealt{2006A&A...445..457M} and Sarria et al., in preparation). 
All but one spectra are typical of optically obscured AGN, thus 
confirming the \xray\ classification (i.e., Type~2).  The only exception is source
PKS~0537\#91, with emission line ratios typical of an HII region (Sarria
et al. in preparation).
One source, Abell~2690\#29, shows the typical rest-frame spectrum of a 
high-redshift dust-reddened quasar, with a broad H$\alpha$ line and a Type 1.9
classification (\citealt{2006A&A...445..457M}).
Spectroscopic redshift $z$ are in the range 0.9-2.08 and are reported in Table~\ref{table1}.

For the sources without redshift (5 out of 16), a photometric redshift was estimated in \cite{2007A&A...468..603P}, where the \spitzer\ data
reduction and a preliminary SED analysis were presented.

Finally, \submm\ observations were performed for four sources in 
October-November 2004. Only one object, GD~158\#19 (z=1.957), was
detected, while for the others upper limits were gained.  
Because of its broad-band coverage (up to 850$\mu$m) with 
good-quality photometric data, the source GD~158\#19 was studied in a
dedicated work (see V09). 

In Table~\ref{table1} we report source name, 2--10 keV flux, R, and K$_{s}$ photometry, X/O ratio, 
the redshift $z$, the column densities N$_H$ and the absorption-corrected 
(2--10 keV) \xray\ luminosities of the sample. The source table order reflects
the {\spitzer} observation strategy (see Sec 3.).
The \submm\ flux densities are reported in Table~\ref{table2} along with 
the fluxes obtained in the IR bands (from 3.6$\mu$m up to 160$\mu$m) 
with the \spitzer\ satellite (see $\S$ \ref{spitzer_par}). 

Almost all the sources have column densities N$_H$ in the range 
$10^{22.0}$--$10^{23.4}$~cm$^{-2}$ and 2--10~keV rest-frame luminosities 
in the range \hbox{10$^{43.8}$--10$^{44.7}$}~\lum, placing them among the Type~2 quasar population.  

\begin{table*}
\begin{center}
\begin{minipage}{18cm}
\caption{Properties of our luminous obscured quasars}
\label{table1}
\begin{tabular}{l  l   c  c  c  c  c  c  c} \\\hline\hline 
Complete Name & Name & 2--10~keV flux    & $R$ & $K_{s}$   & X/O & $z$& $N_H$  & $L_{2-10\ keV}$\\
(1)          &            (2)       &       (3)        &  (4) &    (5)    &
           (6)   & (7) &(8) &(9)            \\\hline    
 HELLAS2XMM 054022.0-283139  & PKS~0537\#43        &     3.35 &    22.70  &   17.50   &1.10   &1.797  &10.5$^{9.4}_{4.8}$&  6.8   \\
HELLAS2XMM 053920.4-283721    & PKS~0537\#11a        &     4.19  &    23.40  &   18.25   &1.48   &0.981  & 1.3$^{1.5}_{0.9}$&  1.9\\
 HELLAS2XMM 053917.1-283819   &PKS~0537\#164       &     1.50  &    23.60  &   19.02   &1.12   & 1.824 & --&3.1 \\
 HELLAS2XMM 053851.3-283949    &PKS~0537\#123       &     2.97  &    23.10  &   17.94   &1.21   & 1.153 & 6.6$^{21.6}_{4.1}$& 2.0\\
  HELLAS2XMM 003413.8-115559       & GD~158\#62          &     3.59  &   23.30   &   21.83&1.38   & 1.568 & 26.3$^{44.7}_{18.1}$& 4.9\\
 HELLAS2XMM 003357.2-120039      &GD~158\#19          &     2.43  &   21.80   &   --      &0.61   & 1.957 & 7.3$^{11.7}_{5.5}$ & 6.3 \\
 HELLAS2XMM 204428.7-105629      &Mrk~509\#01         &     2.10  &   23.85   &   17.88   &1.36   &1.049  & $<$1.1&1.5   \\
 HELLAS2XMM 204349.7-103243      &Mrk~509\#13         &     3.10  &   23.99   &   18.79   &1.59   & 1.261 & 2.5$^{4.6}_{2.2}$ & 2.6\\
 HELLAS2XMM 235956.6-251019  &Abell~2690\#75      &    3.30   &   24.60   &   18.33  &1.85   & 1.3$^{+0.30}_{-0.20}$~$^\dagger$   &15.0$^{20.0}_{8.5}$&3.2 \\
HELLAS2XMM 031343.5-765426    &  PKS~0312\#36   &    1.90  &   24.70   &   19.13  &1.66   & 0.9$^{+0.05}_{-0.15}$~$^\dagger$   &1.0$^{1.2}_{0.9}$&0.7\\
HELLAS2XMM 054021.1-285037   & PKS~0537\#91   &    4.2   &   23.70   &   18.99  &1.60   & 1.538 &45.9$^{102.0}_{36.0}$&8.1\\
 HELLAS2XMM 053945.2-284910  &PKS~0537\#54   &    2.1   &   25.10   &   18.91  &1.86   &$>$1.3~~$^\dagger$    & -- &2.0\\
HELLAS2XMM 053911.4-283717   &PKS~0537\#111  &   2.1 & 24.50 & 17.64 &1.62  & 1.2$^{+0.20}_{-0.10}$~$^\dagger$ &9.1$^{12.4}_{5.3}$&1.7\\
HELLAS2XMM 000111.6-251202  &Abell~2690\#29      &    2.8   &   25.10   &   17.67  &1.99   & 2.08   &2.1$^{2.6}_{1.6}$ & 8.4  \\
HELLAS2XMM 031018.9-765957   & PKS~0312\#45& 2.8 & 24.40 & 18.62  &1.70 & 1.85$^{+0.20}_{-0.30}$~$^\dagger$   & 8.0$^{8.4}_{4.6}$&6.2\\
HELLAS2XMM 005030.7-520046  &BPM~16274\#69       &    2.27   &   24.08   &   17.87  &1.48   & 1.35   & 2.5$^{1.5}_{1.0}$ & 2.4\\\hline\hline
\end{tabular}
\end{minipage}
\vskip0.2cm\hskip2.8cm
\begin{minipage}[h]{18cm}
\footnotesize 
(1) Source complete name; \\
(2) source abbreviated name (adopted throughout the
paper);\\
(3) 2--10 keV \xray\ fluxes in units of $10^{-14}\ $erg\ cm$^{-2}$\ s$^{-1}$ from \cite{2004A&A...421..491P} 
(with the exception of BPM~16274\#69, from Lanzuisi et al., in 
preparation); \\
(4) R-band magnitude from \cite{2003A&A...409...79F} (with the 
exception of source BPM~16274\#69, from \citealt{2007A&A...466...31C});\\ 
(5) K$_{s}$-band magnitudes. For a sub-sample of sources, the 
K$_{s}$-band analysis can be found in \cite{2004A&A...418..827M};\\ 
(6) \xray-to-optical flux ratio;\\ 
(7) source redshifts. Spectroscopic redshift from optical spectroscopy 
(\citealt{2003A&A...409...79F}) and, for three sources, from near-IR spectroscopy (Abell~2690\#29 and BPM~16274\#69 from 
\citealt{2006A&A...445..457M}; PKS~0537\#91 from Sarria et 
al., in preparation); photometric redshifts ($^\dagger$symbol), and corresponding 1$\sigma$ errors, 
from \cite{2007A&A...468..603P};\\
(8) column densities in source rest-frame in units of 10$^{22}$~cm$^{-2}$ measured 
from \xray\ spectral fitting (see \citealt{2004A&A...421..491P} and 
Lanzuisi et al., in preparation). 
For the sources with photometric redshifts, values are taken from 
\cite{2007A&A...468..603P}. For PKS~0537\#164 the \xray\ spectral 
fitting is prevented by the few \xray\ counts; for Mrk~509\#01, only an upper 
limit was derived (\citealt{2004A&A...421..491P}); for 
PKS~0537\#54, the column density is 9.3${\times}10^{22}$ 
cm$^{-2}$ for $z=$ 1.3. Galactic absorption 
column densities adopted in the spectral fitting are: 
$8{\times}10^{20}$~cm$^{-2}$ for the field PKS~0312$-$77; 
$2{\times}10^{20}$~cm$^{-2}$ for the field Abell~2690; $2.1{\times}10^{20}$ 
cm$^{-2}$ for the field PKS~0537$-$28; $4{\times}10^{20}$ 
cm$^{-2}$ for the field Mrk~509; $2.5{\times}10^{20}$~cm$^{-2}$ 
for the field GD~158$-$100 (see \citealt{1992ApJS...79...77S});\\
(9) rest-frame 2--10 keV absorption-corrected \xray\ luminosity in units of 10$^{44}$~erg~s$^{-1}$ from \cite{2004A&A...421..491P} and from \cite{2007A&A...468..603P} for sources with photometric redshifts. 
Luminosities are computed using H$_{0}$=70 Km s$^{-1}$ 
Mpc$^{-1}$, $\Omega_{m}=0.3$ and $\Omega_{\Lambda}=0.7$.\\ 
\end{minipage}
\end{center}
\end{table*}

\section{The Spitzer data}
\label{spitzer_par}
The targets were observed by \spitzer\ in 2006 with both IRAC and MIPS
instruments in photometry mode. All the sources were observed with the same 
total integration time in the IRAC bands (480s), while different strategies
were followed in the MIPS bands taking into account the different optical-NIR properties.  While all the 
objects were observed at 24$\mu$m for a total integration time of 1400s,
only a sub-sample of sources with relatively bright R-band magnitudes 
(R$<$24) and spectroscopic redshifts were observed at longer
wavelengths, with integration times of 300s and
600s at 70 and 160$\mu$m, respectively. 
PKS~0537\#91 (see Table~\ref{table1}) was not observed at 70 and 
160 ${\mu}$m since it had no redshift at the epoch of the \spitzer\ 
observations.

The reduction method is described in detail in
\cite{2007A&A...468..603P} and in V09 and it is briefly
summarized here. The IRAC flux densities of the sources were measured from the
post-basic calibrated data (post-BCD) images in the \spitzer\
archive. Aperture fluxes were measured on the background subtracted
maps within a 2.45$^{\arcsec}$ aperture radius using aperture
corrections of 1.21, 1.23,1.38 and 1.58 for the four IRAC bands
(following the IRAC Data Handbook). For the MIPS bands, we started the
analysis from the basic calibrated data (BCD) at 24$\mu$m and from
the median high-pass filtered BCD (fBCD) at 70 and 160$\mu$m, as
suggested for faint sources.  At 24$\mu$m, the BCD were corrected for a
residual flat fielding dependent on the scan mirror position (see
\citealt{2006AJ....131.2859F}; \citealt{2007A&A...468..603P}). 
We constructed then our own mosaics using
the SSC MOPEX software (\citealt{2005PASP..117.1113M}). 
Aperture fluxes were measured
within a 7$^{\arcsec}$ aperture radius for the 24$\mu$m band and
16$^{\arcsec}$ aperture radius for the 70 and 160$\mu$m band. The aperture
corrections used were 1.61, 2.07 and 4.1, respectively (see the
MIPS Data Handbook). A small aperture radius was used at longer
wavelengths (at 160$\mu$m the adopted radius is comparable to half of
the PSF FWHM) to exclude the contamination of nearby far-infrared
sources (see V09). At 24$\mu$m, thanks to a
better PSF sampling, two sources (PKS~0312\#36 and Abell~2690\#29) were
deblended using a PSF deconvolution analysis.

All  16 sources were clearly detected in the IRAC bands. At 24$\mu$m, 
14 sources (out of 16) were detected above the 5$\sigma$ level 
and span almost two orders of magnitude in flux, from $\sim$7000~$\mu$Jy 
down to the faintest source, close to the 5$\sigma$ detection 
level ($\sim$100~$\mu$Jy). For the two sources without detection, 
an upper limit (3$\sigma$) was estimated from the average noise of 
the map, derived by making multiple aperture
measurements at random locations throughout the residual mosaic after
source extraction. The typical average noise (1$\sigma$) 
is 20~$\mu$Jy. At 70 and 160$\mu$m, as said before, only the brightest 
R-band sources were observed; among
them, only the two most luminous (in the optical band) were detected, 
PKS~0537\#43 and GD~158\#19, the latter being the source described in V09. 
For the remaining 6 sources, an upper limit (3$\sigma$) was estimated from the residual mosaic (see
also \citealt{2006ApJ...647L...9F}) after
source extraction, as done at 24$\mu$m. The typical average noise (1$\sigma$) 
is 1.2~mJy and 8~mJy at 70 and 160$\mu$m, respectively (consistent with
the results obtained in the COSMOS field from \citealt{2009arXiv0902.3273F} 
taking into account the different integration time).

Table~\ref{table2} reports the target flux densities provided by 
\spitzer. To compute uncertainties, the noise map was added in 
quadrature to the systematic uncertainties, assumed to be 10 per cent 
in the IRAC and MIPS 24$\mu$m bands and 15 per cent at 70 and 160$\mu$m 
(see IRAC and MIPS Data Handbook).

\begin{figure}
\includegraphics[angle=-270,width=0.5\textwidth]{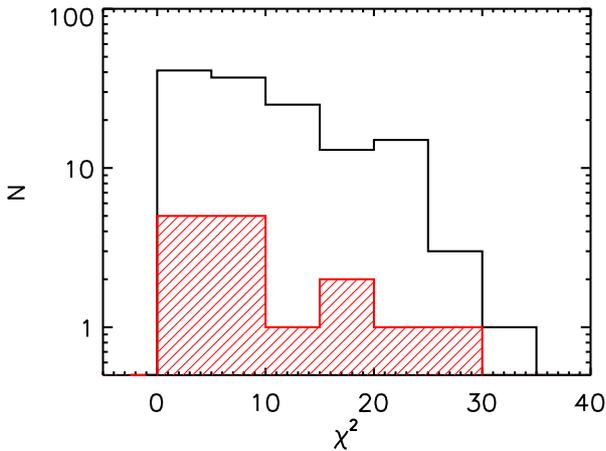}
\caption{$\chi^{2}$ distribution. The red hatched and the empty
  distributions represent the best-fitting solutions and all the
  solutions at 1$\sigma$, respectively. PKS~0537\#43 is not
  reported given the high $\chi^{2}$ value (Table~\ref{table3}).}
\label{figure_chi}
\end{figure}


\begin{table*}
\begin{minipage}{\textwidth}
\caption{\spitzer\ and  \scuba~ flux densities}
\label{table2}
\centering
\begin{tabular}{l  r   r   r   r   r   r  r  r  r} \\\hline\hline 
Source name & 3.6$\mu$m    &   4.5$\mu$m  & 5.8$\mu$m &  8.0$\mu$m
           &  24$\mu$m &  70$\mu$m &  160$\mu$m   &  450$\mu$m   &  850$\mu$m     \\
           & [$\mu$Jy]  & [$\mu$Jy]    & [$\mu$Jy] &  [$\mu$Jy] &  [$\mu$Jy] & [mJy] &  [mJy]  & [mJy]   &  [mJy]  \\\hline
      
PKS~0537\#43         &   423 $\pm$   42&  918  $\pm$  92&  1750$\pm$175 &     3202  $\pm$ 320	&  6879	$\pm$688& 27.0$\pm$4.5 &  39.6$\pm$  7.8 &      &  \\
PKS~0537\#11a         &   57  $\pm$   6 &   69  $\pm$  7 &    95$\pm$ 10 &      168  $\pm$  17	&   517	$\pm$57 &   $<$3.6 &  $<$24  &      &   \\
 PKS~0537\#164        &   21  $\pm$  2  &   18  $\pm$  2 &    19$\pm$ 4 &      23  $\pm$   5	&   $<$60   &   $<$3.6 &  $<$24  &      &   \\
 PKS~0537\#123        &   84  $\pm$   9 &   96  $\pm$ 10 &   113$\pm$ 13 &      165  $\pm$  17	&   747	$\pm$81 &   $<$3.6 &  $<$24  &      &   \\
 GD~158\#62           &   60  $\pm$   6 &   91  $\pm$  9 &   155$\pm$ 17 &      339  $\pm$  35	&  1608	$\pm$164&   $<$3.6 &  $<$24  &   $<$74   & $<$6.1 \\ 
 GD~158\#19           &  226  $\pm$  23 &  387  $\pm$ 39 &   756$\pm$
 76 &      1547 $\pm$ 155	&  5326	$\pm$534& 27.0$\pm$4.5&37.7$\pm$ 7.8&$<$94.3 & 8.6$\pm$2.1\\

 Mrk~509\#01          &   63  $\pm$   7 &   51  $\pm$  6 &    60$\pm$
 11 &      59  $\pm$  10	&  $<$60	 &    $<$3.6  & $<$24
 &     $<$82   &  $<$6.6  \\  
 Mrk~509\#13          &   51  $\pm$  6  &   69  $\pm$  7 &   110$\pm$ 11 &      215  $\pm$  17	&   866	$\pm$ 94&    $<$3.6  & $<$24 &     $<$75  &   $<$6.2	    \\
 Abell~2690\#75       &   51  $\pm$  5  &   56  $\pm$  6 &    89$\pm$ 11 &      139  $\pm$  15	&   565	$\pm$ 62&    &         &     & 	    \\
 PKS~0312\#36    &   41  $\pm$  4  &   44  $\pm$  5 &    40$\pm$  8 &      710  $\pm$  9	&   236	$\pm$ 30&    &       &     & 	    \\
 PKS~0537\#91    &   28  $\pm$  4  &   35  $\pm$  4 &    42$\pm$  8 &      800  $\pm$  10	&   301	$\pm$ 40&     &       &    & 	   \\
 PKS~0537\#54    &   31  $\pm$  4  &   35  $\pm$  4 &    50$\pm$ 10 &      470  $\pm$  8	&  279	$\pm$ 45&       &       &   &	    \\
 PKS~0537\#111   &   88  $\pm$  9  &   75  $\pm$ 8	 &    41$\pm$  6 &      460  $\pm$  7	&  148	$\pm$ 28&        &      &     & 	    \\
 Abell~2690\#29       &   141 $\pm$  14 &  185  $\pm$ 19 &   260$\pm$ 27 &      371  $\pm$  38	& 1012	$\pm$ 106&      &      &     & 	    \\
 PKS~0312\#45    &   50  $\pm$   6 &   62  $\pm$  7 &    69$\pm$ 10 &      780  $\pm$  10	&  249	$\pm$  35&      &       &     & 	    \\
 BPM~16274\#69        &   86  $\pm$   9 &   92  $\pm$  9 &    97$\pm$ 11 &      120  $\pm$  13	&  286	$\pm$  34&      &       &    & 	   \\
\hline\hline
\end{tabular}
\end{minipage}
\vskip0.2cm\hskip0.4cm
\begin{minipage}[h]{17.8cm}
\footnotesize
The upper limits are given at the 3$\sigma$ confidence level.\\
\end{minipage}
\end{table*}

\section{Modeling the spectral energy distribution}
\label{model_sec}

 The observed optical-to-MIR (or FIR/sub-mm) SEDs can be modelled as the
 sum of three distinct components:  a stellar component, which emits most of
 its power in the optical/NIR, an AGN component, whose emission peaks in
 the MIR for obscured quasars, and is due to hot dust
  heated by UV/optical radiation from gas accreting onto the central SMBH, 
 and a starburst component, which represents the major contribution to the FIR spectrum.

In this work, we considered all the three components (see Sec. 4.1, 4.2 and
\citealt{2008MNRAS.386.1252H}).  Since the focus of the paper is on the AGN
contribution to the SED, we discuss in some more detail the hot dust 
modeling and its uncertainties. 
 
The hot dust emission in AGN is reproduced using the F06 model. This model
follows the formalism developed by different authors (e.g., \citealt{1992ApJ...401...99P}; 
\citealt{1994MNRAS.268..235G}; \citealt{1995MNRAS.273..649E}), where 
the IR emission in AGN originates from dusty gas around the SMBH with 
a smooth distribution. 
The dust grains are heated by high-energy photons coming
from the accretion disk, and their thermal and
scattering re-emission, mostly at IR wavelengths, is computed by means of
the radiative transfer equations.  For what concerns the dust
distribution geometry, different possibilities  (i.e. ``classical'' torus
shape, tapered or flared disk), are explored in the literature. 

More recently, models considering a clumpy distribution for the dust 
have been developed  (e.g., \citealt{2002ApJ...570L...9N}; 
\citealt{2006A&A...452..459H}; \citealt{2008ApJ...685..147N}). These models
successfully explain many recent observations in the mid-IR, such as the strength
of absorption and emission features at 9.7 ${\mu}$m and the \xray\ variability
(\citealt{2002ApJ...571..234R}).

A further possibility to the torus models described above are the 
disk-wind models (see \citealt{2006ApJ...648L.101E} and references therein),
involving a completely different 
approach. The dusty clouds, responsible 
for the obscuration, are part of an hydro-magnetic wind coming
from the accretion disk. On the one hand, these models are potentially 
capable of explaining disparate phenomena in AGN 
(from broad emission to absorption lines and obscuration), 
providing an hydro-dynamical justification for the persistence 
of the clouds around the SMBH. 
On the other hand, a parameterization which takes into account the 
observational constraints on the clumpy obscuration, hence supplying a 
grid of synthetic IR SEDs, does not exist yet 
(see \citealt{2008NewAR..52..274E}).

A detailed comparison between smooth and clumpy dust distribution models is 
discussed by \cite{2005A&A...436...47D}. They conclude that 
both models yield similar SEDs (see also \citealt{2008NewAR..52..274E} and 
\citealt{2008ApJ...685..147N}). 
The main difference is in the strength of the silicate feature observed in
absorption in objects seen edge-on, which is, on average, weaker for clumpy
models with the same global torus parameters. In clumpy models, in fact, clouds at different distances from the central source
can be intercepted by the line of sight, including the innermost clouds,
where the silicate feature is in emission given the higher temperature of the dust
grains.

A systematic comparison of the two model predictions is beyond the scope of
the present paper and should be performed on
high-quality IR data (i.e. a \spitzer\ {\it IRS} spectroscopic sample). 

Notwithstanding these limitations, with the present work we aim at 
extracting the maximum information using the available photometric data. 
The F06 model adopted in this work is one of the models best tested 
against both broad-band photometry (F06; 
\citealt{2004A&A...418..913B}; \citealt{2007MNRAS.376..416R}; 
\citealt{2008MNRAS.386.1252H, 2009MNRAS.tmp.1247H}; 
\citealt{2009ApJ...697.1010A}; V09) and Spitzer mid-infrared 
spectra (F06). Moreover, the F06 model was the first able to reproduce the 
quasar mid-IR spectra, considered a very 
important constraint to characterize the dust properties in AGN and probe 
the Unified Model.

\begin{figure*}[t]
\centering
\includegraphics[width=0.9\textwidth]{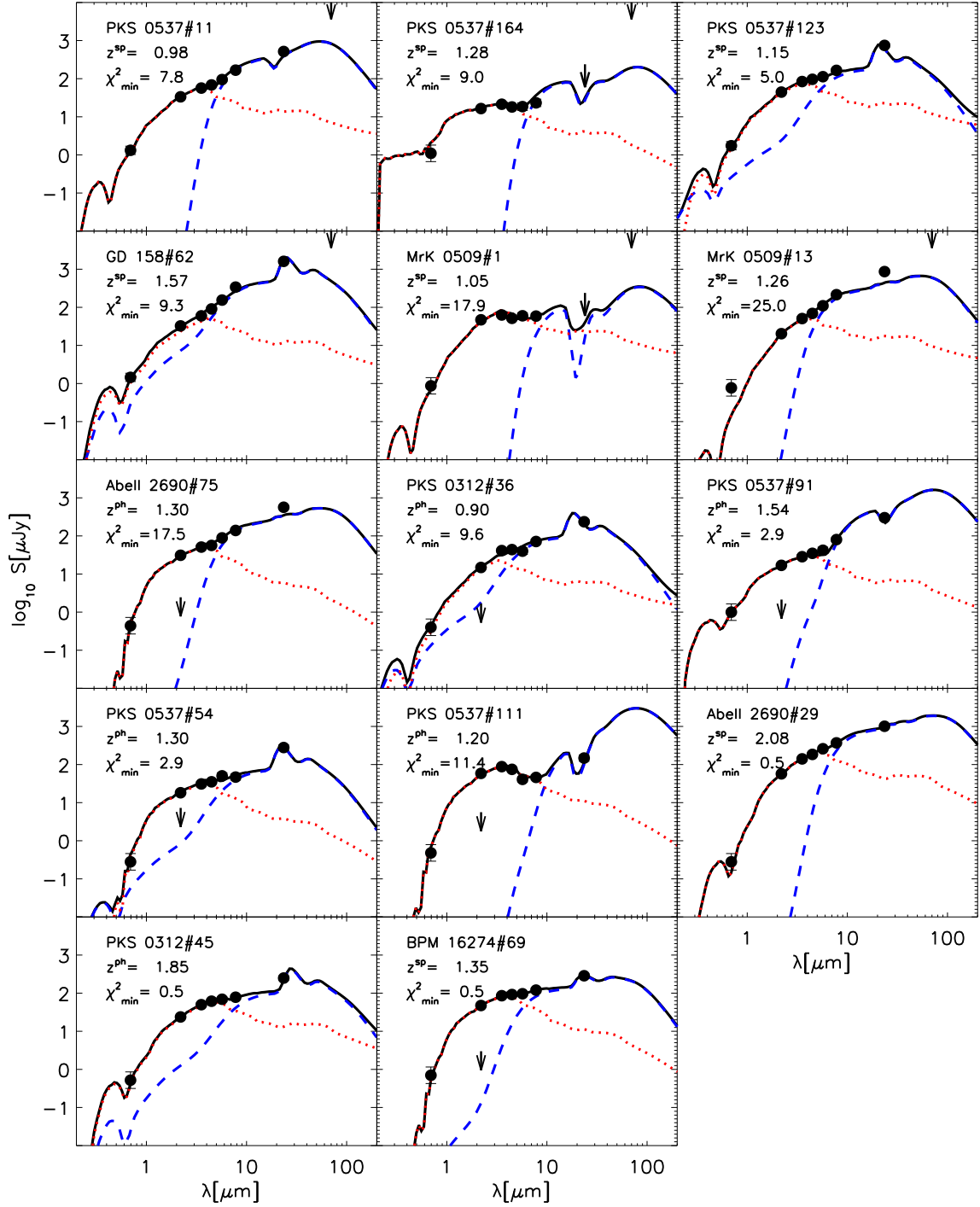}
\caption{{\bf a.} Observed-frame SEDs for 14 sources with data from the 
R-band to the 24$\mu$m (black dots) compared with the best-fit model 
obtained as the sum (solid black line) of a stellar component 
(red dotted line) and an AGN component (blue dashed line). The nuclear K$_{s}$-band upper limits (downward-pointing arrows) were derived 
from the morphological analysis carried out by \citet{2004A&A...418..827M}.} 
\label{figure_sed}
\end{figure*}

\addtocounter{figure}{-1}
      
\begin{figure*}
\hskip3.6cm
\includegraphics[width=17cm]{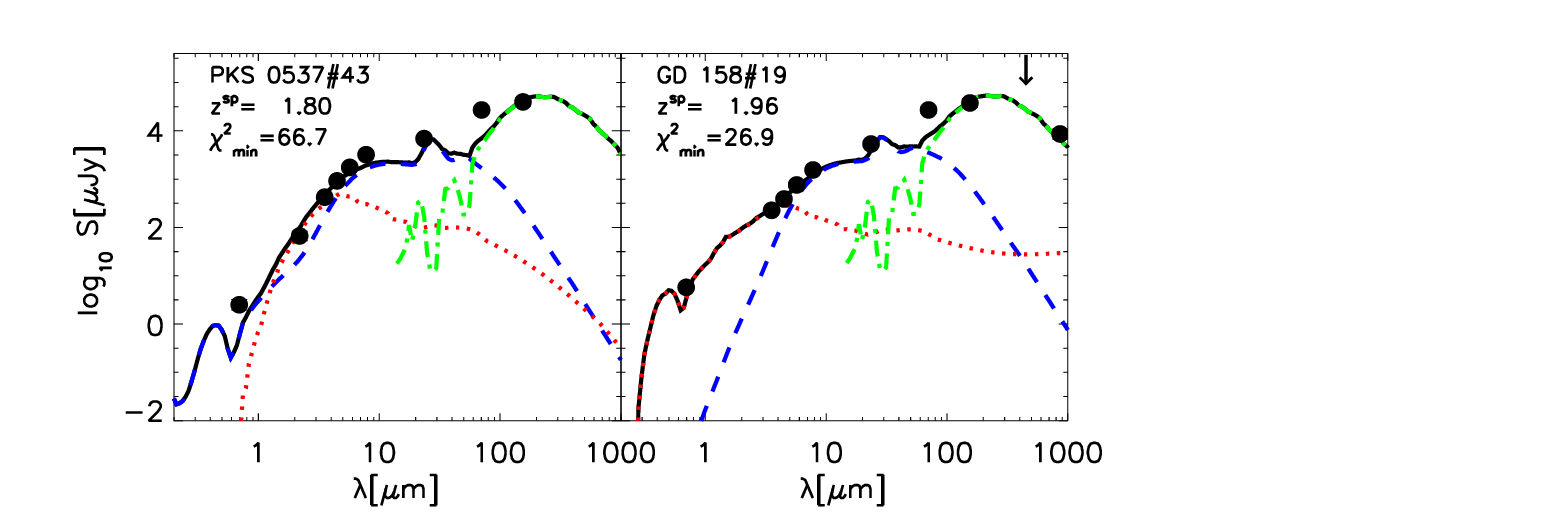}
\caption{{\bf b.} As in Fig.~\ref{figure_sed}a for sources with detections 
also at longer wavelengths (FIR/\submm). An additional starburst component 
(green dot-dashed line) is considered in the best-fit model.}
\end{figure*}

\subsection{The AGN-torus component}
\label{torus_comp_sec}

The F06  code assumes a smooth dust distribution around the central source 
consisting of a Galactic mixture of silicate and graphite grains.
The presence of silicate dust grains is clear from the absorption feature at 
9.7 ${\mu}$m seen in most of Type~2 AGN. The graphite grains are, instead, responsible 
for the rapid decline of the emission at wavelengths shortwards 
of a few microns, corresponding to a blackbody emission of about 1500 K, the
sublimation temperature of these grains (see F06).

The assumed dust geometry is a flared disk 
(see \citealt{1995MNRAS.273..649E}), that is a sphere with the polar cones 
removed. The internal radius of the dust distribution is defined by the 
sublimation temperature of the dust itself. In order to simulate a more 
realistic shape for the dust distribution, F06 assumes that the dust density 
can vary both with the radial and the angular coordinates:

\begin{equation}
\label{eq_density}
\rho(r,\theta)=\rho_0\cdot r^{\alpha}\cdot e^{-\gamma\cdot |\cos(\theta)|}
\end{equation}

\noindent  where $\theta$ is the angle with respect to the equatorial plane.


The dusty torus is heated by the emission of the inner accretion disk, 
which represents the input energy to the radiative transfer code. 
The assumed spectrum for the accretion disk is defined in the 10$^{-3}$ to
20${\mu}$m regime (from soft X-rays, i.e. 1.25 keV, to mid-IR) and is
parameterized by broken power laws in F06. The specific indices of the power laws are adapted from the Granato \& Danese (1994) 
and the Nenkova (2002) models and are consistent with the broad-band SEDs  of a sample of Type~1 AGN from the SDSS survey 
(\citealt{2008MNRAS.386.1252H}; see their Figs.~8 and 9).

Along with the thermally re-processed light, the F06
provides, as a function of the line-of-sight and optical depth, the
fraction of the inner accretion disk light not intercepted by the torus and
the scattered light. 
In the following, with AGN component we will refer to the sum of all the three contributions. 

\subsection{The stellar and starburst components}
\label{star_comp_sec}

The stellar component is modelled as the sum of Simple Stellar 
Populations (SSP) models of different age, all assumed to have common (solar) 
metallicity. A \cite{1955VA......1..283S}
initial mass function (IMF) with mass in the range (0.15-120
M$_{\odot}$) is assumed. The SSP spectra have
been weighted by a Schmidt-like law of star
formation (see \citealt{2004A&A...418..913B}):


\begin {equation}
SFR(t)=\frac{T_{G}-t}{T_{G}}{\times}\exp\left({-\frac{T_{G}-1}  {T_{G}{\tau}_{sf} }}\right)  
\end{equation}

\noindent where $T_{G}$ is the age of the galaxy (i.e. of the oldest SSP) 
and $\tau_{sf}$  is the duration of the burst in units of the
oldest SSP (see V09). As in \cite{2008MNRAS.386.1252H} and V09, a common 
value of extinction is applied to stars at all ages, adopting the extinction 
law of our Galaxy  ($R_{V}=3.1$; \citealt{1989ApJ...345..245C}).  

In order to keep the number of free parameters 
as low as possible, emission from cold dust, which dominates the bolometric emission at 
wavelengths longer than 30${\mu}$m rest-frame, is included only 
when far-IR/\submm\ data allow us to constrain that part of the SED 
(two sources of the sample). For the same reason, additional components, such as the cold
absorber detached from the torus (i.e. \citealt{2008ApJ...675..960P}), 
which might improve the fit but would increase the complexity of the 
overall modeling, is not included. 
To reproduce the starburst component, a set of semi-empirical models of 
well known and studied starburst galaxies is used, similarly to V09.

\subsection{SED fitting procedure}
\label{fitting_sec}

The quality of the fitting solutions is measured using a 
standard $\chi^{2}$ minimization technique (as also done in 
\citealt{2008MNRAS.386.1252H}), where the observed values are the 
photometric flux densities (from optical-to-MIR/FIR) and the model values 
are the ``synthetic'' flux densities obtained by convolving the sum of stars, 
AGN and starburst components through the filter response curves. 

%
%

Before starting the general fitting procedure, we test which torus 
parameters mainly influence the global model SED and are more sensitive to 
our data sets. Parameters which are not constrained by our data were 
then frozen.

The F06 torus model is described by six parameters: the ratio
$R_{max}/R_{min}$ between
the outer and the inner radius of the
torus (the inner radius being defined by the sublimation temperature of the 
dust grains); the torus full opening angle $\Theta$; the optical depth
$\tau$ at 9.7$\mu$m ($\tau_{9.7})$; the line of sight $\theta$ with
respect to the
equatorial plane, and two parameters, $\gamma$ and $\alpha$, describing the 
law for the spatial distribution of the dust and gas density $\rho$ inside the torus (Eq. \ref{eq_density}).


In our approach, we let free to vary, inside the
pre-constructed grid of torus models the following parameters: the torus full
opening angle $\Theta$, the optical depth
$\tau_{9.7}$ and the parameter $\alpha$ describing the radial
dependence of the density. We fix
$R_{max}/R_{min}$=30, which translates into compact tori of a few tens of 
parsecs (given that $R_{min}$ is directly connected to the sublimation
temperature and to the accretion luminosity of the central
BH). 
 Recent high-resolution IR observations support a compact dust distribution in nearby luminous AGN. 
Using the interferometry at VLTI in the 8-13 ${\mu}$m band, a torus of size ${\sim}$2-3 pc was detected in NGC~1068 (\citealt{2004Natur.429...47J}). Similar compact tori were also found in 
other local AGN, as Circinus and NGC~4151 (see the review by 
\citealt{2008NewAR..52..274E}).

Regarding the density
distribution, we allow power-law profiles decreasing with the radius
with different coefficients $\alpha$. We do not allow any dependence on the distance from the equatorial plane by fixing $\gamma$=0. 
As a result, different angles for the lines of sight $\theta$ (with respect 
to the equatorial plane) give the same SED, once the torus is intercepted.

Given the F06 grid of models, the discrete values, allowed for our free 
parameters, are:
$\Theta = [60, 100, 140^{\circ}] $, $\tau_{9.7} = [0.1, 0.3, 0.6, 1, 2, 3, 6, 10] $ and $\alpha =[ -1, -0.5, 0]$, implying 72 different torus SEDs.

Considering the stellar component,  we fix at $z=4$ the redshift
for the formation of the oldest SSP, i.e.,  given the observed redshift of the
sources, we consider galaxies with ages typical of early-type galaxies
(${\gsimeq}$1-2 Gyr). This assumption is justified by the observed R-K$_{s}$
colours and the brightness profiles typical of early-type galaxies as obtained
by a detailed morphological analysis in the K$_{s}$-band
(\citealt{2004A&A...418..827M}). Concerning the star-formation history, we allow the $\tau_{sf}$ parameter 
of the Schmidt-like law and the value of the 
extinction $E(B-V)$ to vary; the latter is a key parameter, along with 
the optical depth $\tau_{9.7}$ of the torus, in shaping the
optical-NIR continuum.

Overall, the SED-fitting procedure ends with 5 free parameters (6 when
also a starburst component is included). 
Since the problem is
affected by some degeneracy, we consider, along with the best-fitting
 solutions, all the acceptable solutions within
 1$\sigma$ confidence level (by considering, for
each source, all the solutions with
$\chi^{2}-\chi^{2}_{min}=\Delta\chi^{2}{\lsimeq}$5.89 or 7.04, when
the cold starburst component is added, see \citealt{1976ApJ...208..177L}).

\section{Results from SED fitting}
\label{results_fitting}

In Fig.~\ref{figure_chi} we show the $\chi^{2}$ distribution. 
The hatched histogram represents the distribution of
the best-fitting solutions for our 16 targets, while the empty histogram 
shows the $\chi^{2}$
distribution of all the solutions within 1$\sigma$, satisfying the criteria 
$\chi^{2}{\lsimeq}\chi^{2}_{min}$+$\Delta\chi$. Given our adopted grid
for the fitted parameters, the total number of
solutions at 1$\sigma$ (constructed by adding all the solutions at
1$\sigma$ of each object) is 137, 8 solutions on average per source
(including the best-fitting one).

The two distributions do not show a significant difference, and this 
reinforces our choice of considering, in the analysis of the
parameter space and degeneracy, all the solutions at 1$\sigma$
as a unique statistical
sample.

In terms of the absolute values of the $\chi^{2}$, only 7 sources (out of
16) give a formally acceptable fit
($P(\chi^{2}>\chi^{2}_{obs})>90\%$, see Table \ref{table3}); the remaining sources have a best-fit 
model with large $\chi^{2}_{obs}$. While we use the $\chi^{2}$ to assign 
a relative goodness of different parameter combinations inside the parameter 
grid, we will not take the absolute probabilities
at face value. Over-estimated  $\chi^{2}$ are, in fact, a common
problem of most SED fitting techniques, resulting from a combination of two different reasons: the limited grid of models (72 torus
models with the adopted choice of parameters, see $\S$\ref{fitting_sec}) 
with no uncertainties associated, and the photometric measurements 
with often underestimated uncertainties (see \citealt{2008ApJ...684..136G} 
for a detailed description of this issue). 

In Fig.~\ref{figure_sed}a,b the observed SEDs, from the R-band to the IR 
(or \submm), are reported with the best-fitting models over-plotted 
(solid line). All the sources need a host galaxy component (red dotted line) 
and an AGN one (blue long-dashed line). 
The stellar component
dominates in the R and K$_{s}$ bands, while the nuclear one at 24$\mu$m. 
In the IRAC bands, both components contribute, the fraction
depending on the properties of the individual sources. For PKS~0537$\_$43 and G158$\_$19, where data points at longer wavelengths 
are available, an additional starburst component is needed
(green dot-dashed line in Fig.~\ref{figure_sed}b).

\begin{figure}
\centering
\includegraphics[width=7cm]{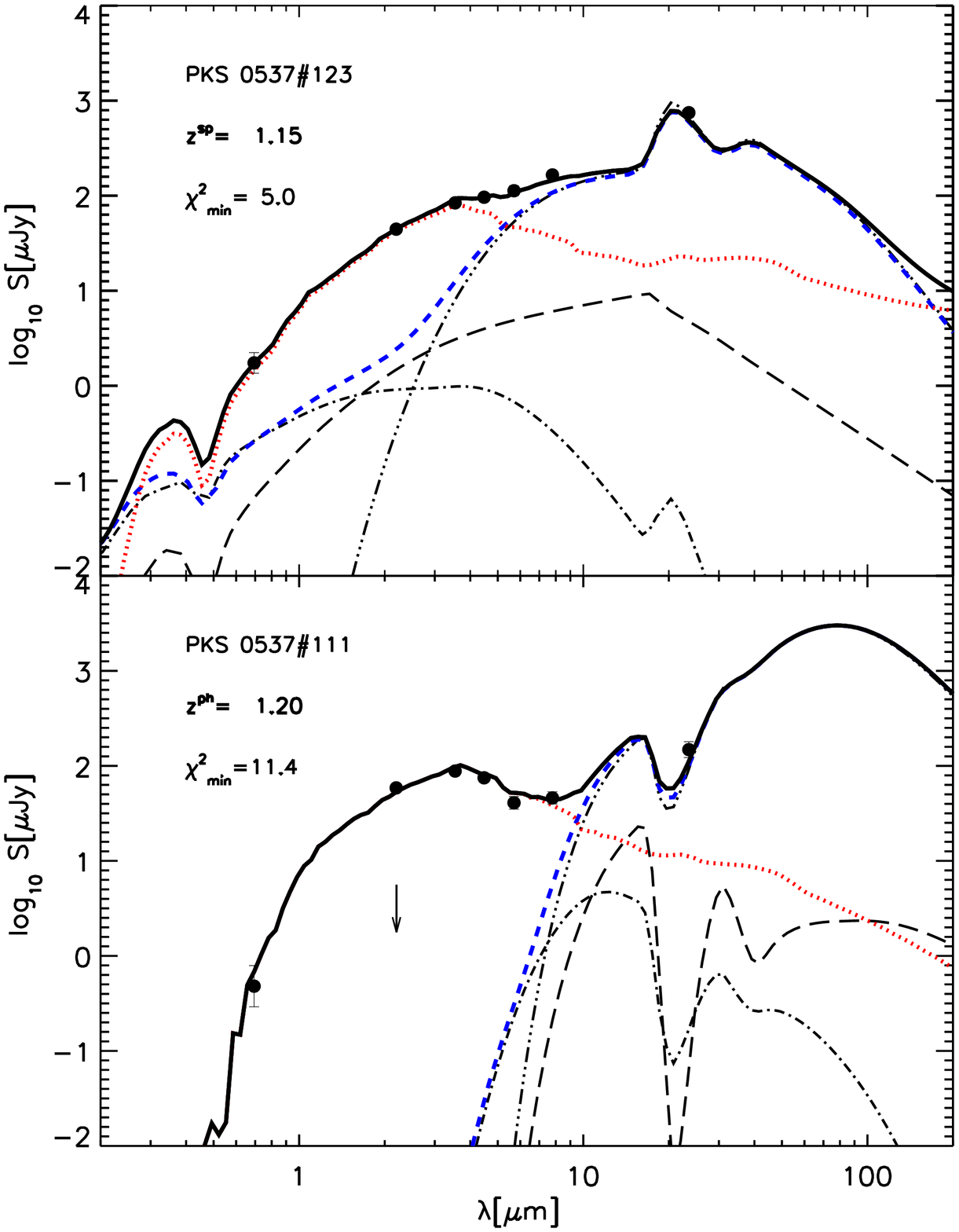}
\caption{Observed-frame spectral energy distribution as in Figs. 3a,b for 2
  sources characterized by extreme values of
  the optical depth: $\tau_{9.7{\mu}m}$=0.1 and 10 for PKS~0537$\_$123 and PKS~0537$\_$111 respectively. In this Figure, the 3 different components that contribute to the AGN emission are reported: direct nuclear light
  (long-dashed line), scattered light (dashed-dot line) and thermally
  re-emitted light (dashed-dot-dot-dot line).} 
\label{figure_sed_agn}
\end{figure}

 In Fig.~\ref{figure_sed_agn}, the relative contributions of the thermal, direct
and scattered light to the total AGN light are shown for two sources characterized by a low
($\tau_{9.7}$=0.1, PKS~0537\#123) and a high ($\tau_{9.7}$=10, PKS~0537\#111) optical depth, respectively. 
While for $\tau_{9.7}$=10, the AGN emission is dominated by the reprocessed emission 
in all the UV/optical/IR bands, for  $\tau_{9.7}=$0.1 the direct and scattered
components account for the optical/UV AGN emission. Nevertheless, the contribution 
of the components mentioned above never exceeds the 20\% of the observed flux
in the R-band.  
For a sample of highly polarized red active galactic nuclei
selected from the 2MASS survey, a larger contribution of the scattered nuclear
component to the
optical and near-IR emission was found (Cutri et al. 2002,
\citealt{2009ApJ...692.1143K}). The different result obtained from our analysis is probably due to the different degree of
obscuration of the two samples: the 2MASS sample is characterized by column
densities around 10$^{22}$ cm$^{-2}$, while the present sample has a median
column density of $7{\times}10^{22}$ cm$^{-2}$ .

The negligible contribution of the AGN component, relative to the
stellar one, at short wavelengths, is consistent with the upper limits to the
AGN emission derived by
\cite{2004A&A...418..827M} from the analysis of the K$_{s}$-band images (shown
as downward-pointing arrows in Figs. \ref{figure_sed} and \ref{figure_sed_agn}).

In Table~\ref{table3}, the $\chi^2$ values (and the corresponding degree of
freedom)  of the best-fitting solutions are reported for each source.


\subsection{Torus parameters}
\label{torus_fitting}

In the following section, we discuss how the model torus parameters are 
constrained by our data set. As anticipated, we will consider all the 137 
solutions at the 1$\sigma$ level.

First of all, the torus is seen almost edge-on in all the solutions 
(i.e., the line of sight always intercepts the obscuring material), 
in agreement with the Type~2 \xray\ and optical classification of our sources.

The torus model parameters which are left free to vary within the grid of 
models are: the torus opening angle $\Theta$,  the slope $\alpha$ of the dust 
density profile and the optical depth $\tau_{9.7}$. 

By converting the torus opening angle into a covering factor (CF) representing
the fraction of solid angle covered by the dusty material, we find that
solutions with high and low CF are possible, with a slight preference towards
tori with large CF; the mean CF value is 0.65 (1$\sigma{\sim}0.25$), 
corresponding to a torus opening angle of ${\sim}110^{\circ}$. 

According to \cite{2007A&A...468..979M}, the covering factor of the
circum-nuclear dust decreases for increasing optical luminosity at 5100~\AA\ ($L_{5100}$). This relation is explained in terms of a 
``receding-torus''. In \cite{2007A&A...468..979M} the luminosities at $5100$~\AA\ were derived from optical spectroscopy and the CF values from the ratios 
between the 6.7${\mu}$m and the 5100~\AA\ luminosities for a sample of Type~1 quasars spanning five orders of
magnitude in optical luminosity. In our approach, $L_{5100}$ is estimated, 
for each solution, from the input accretion-disk spectrum 
(see Sec. \ref{torus_comp_sec}), once the normalization is found 
(see Sec. \ref{bh_luminosity}). 
The average value of CF and $L_{5100}$ for our sample lies, within 1$\sigma$,
on the relation found by \cite{2007A&A...468..979M}. 
Unfortunately, given the limited range of 
$L_{5100}$, we cannot investigate the 
validity of the CF vs. $L_{5100}$ relation over the range probed by Maiolino
et al. (2007). 

Regarding the density profile, about $65\%$ of the solutions have $\alpha=0$, while $\alpha=-0.5$ and $\alpha=-1.0$ represent $20\%$ and 
$15\%$ of the solutions, respectively. This is reflected also in the 16 
best-fitting solutions, where only 2 sources (Mrk~0509\#1 and PKS~0537\#111) 
are fitted with $\alpha=-0.5$, one with $\alpha=-1.0$ (PKS~0537\#43), 
and the remaining ones with $\alpha=0$ (see Table~\ref{table3}).

\begin{figure}
\centering
\includegraphics[angle=-270,width=7cm]{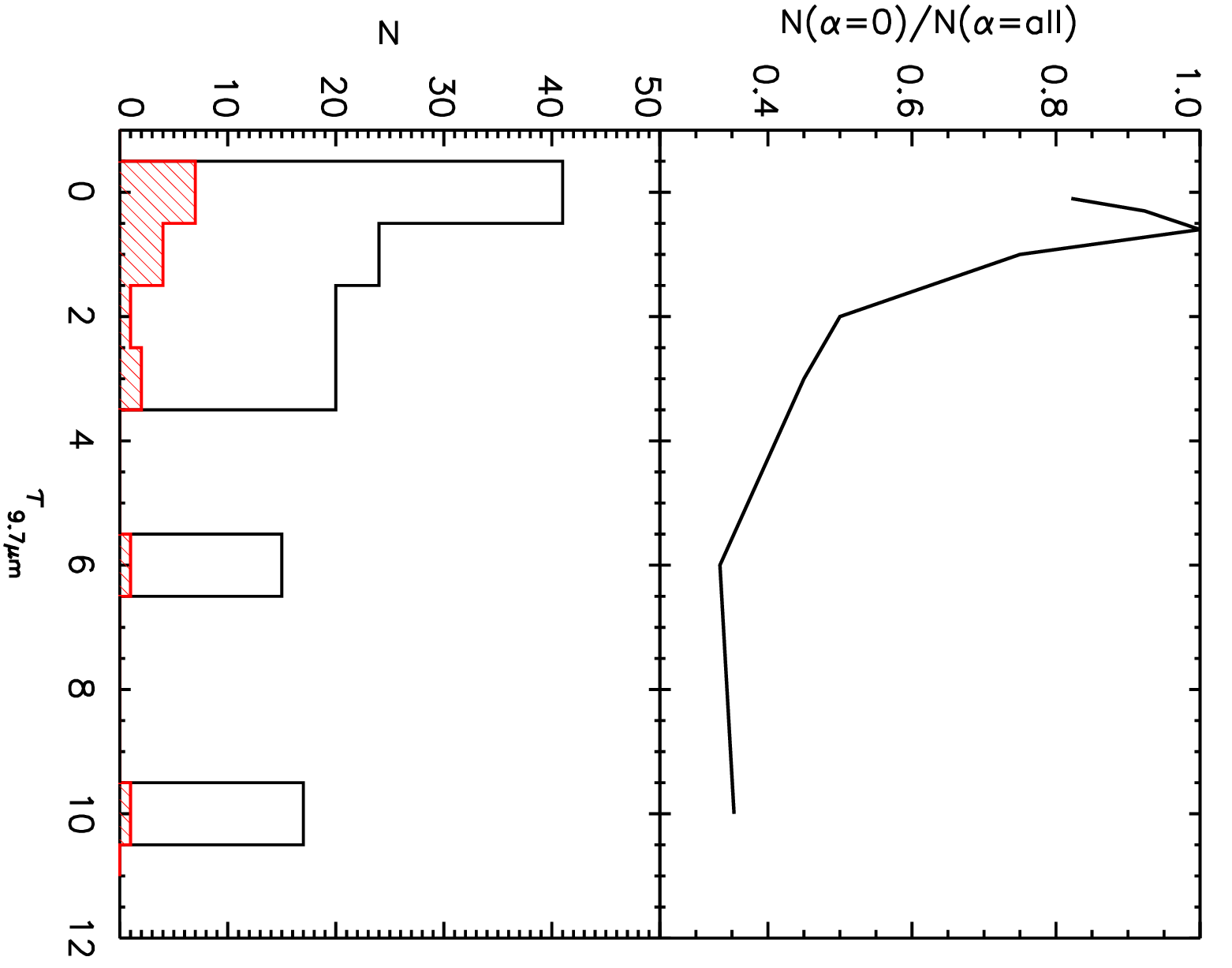}
\vspace{0.5cm}
\caption{$Top$ panel: the fraction of solutions with flat density profile 
($\alpha=0$) as a function of $\tau_{9.7}$. $Bottom$ panel: 
the $\tau_{9.7}$ distribution. The hatched and the empty 
distributions represent the best-fitting (16) and all the solutions (137) 
at 1$\sigma$, respectively.}
\label{figure_tau}
\end{figure}

Solutions with a moderate optical depth $\tau_{9.7}$ are favoured by the
SED-fitting analysis. As shown in Fig.~\ref{figure_tau} 
($bottom$ panel), there is a small number of solutions with high optical 
depths while the majority of the solutions (${\sim}80\%$) are characterized 
by `moderate' $\tau_{9.7}$ ($\tau_{9.7}{\le}3$) and 50$\%$ by
low $\tau_{9.7}$ ($\tau_{9.7}{\le}1$). 
The median value for $\tau_{9.7}$ is 2.  

The finding of a preferred range of optical depths by the SED-fitting, even with relatively sparse photometric 
data, comes from the overall shape of the NIR/MIR continuum. In fact, once the stellar component is determined by the optical/NIR data, the slope 
of the torus component is directly linked to the amount of absorption (i.e., to 
the optical depth) and is relatively well constrained by the available data.

As shown in Fig.~\ref{figure_sed}a,b, for the very low values of the optical depth $\tau_{9.7}$  ($\tau_{9.7}$=0.1), the
F06 model predicts spectra with a weak emission line at 9.7 ${\mu}$m.

We clearly find that the optical depth $\tau_{9.7}$ and the density profile 
$\alpha$ are not independent parameters, since low optical depth solutions 
mostly occur with flat density profile ($\alpha$=0). This is shown in 
Fig.~\ref{figure_tau} ($top$ panel), where the fraction of solutions with
$\alpha$=0 is reported as a function of $\tau_{9.7}$. For the assumed flared
disk geometry, at high optical depth a flat density profile produce too much
IR emission due to the large amount of dust at high radii. Thus the two best-fitting 
solutions with the highest $\tau_{9.7}$ found ($\tau_{9.7}$=6 and 10 
for Mrk~509\#01 and PKS~0537\#111 respectively,
see Fig.~\ref{figure_tau} and Table~\ref{table3}) have a density
profile decreasing with the distance from the central BH ($\alpha$=-0.5).

Recalling that in our procedure different lines of sight are equivalent 
(having assumed $\gamma=0$ in the radial density profile, Eq. 1), we can  convert the optical depths to column densities N$_H$ (adopting a Galactic
dust-to-gas ratio) to be compared with the N$_H$ derived from the \xray\ 
observations. 
Despite the uncertainties affecting the derivation of N$_H$ from the
dust optical depths 
(i.e., dust and gas spatial distributions could be different), 
as well as those affecting the N$_H$ values from X-rays 
(see \citealt{2004A&A...421..491P} for details and Table 1), the two independent estimates 
give a consistent picture for the
majority of the sources, once the 1$\sigma$ uncertainties,
derived from the SED and \xray\ fitting procedure, are taken into
account. By excluding the two sources without a measured N$_H$ from
the X-rays analysis (see Table~\ref{table1}), the median values for N$_H$ are
${\sim}7{\times}10^{22}$~cm$^{-2}$ and ${\sim}$5.5${\times}10^{22}$~cm$^{-2}$, 
from the X-rays analysis and the dust optical
depths, respectively. Therefore, the SED-fitting method confirms the
\xray\ classification of the sources as moderately obscured Compton-thin AGN.

Two sources have a significantly different N$_H$ (by an order of magnitude) 
derived from the two methods, PKS~0537\#111 and Mrk~509\#01. 
These objects are those characterized  by the highest optical depths 
($\tau_{9.7}=6,10$, which are converted into 
N$_H{\sim}5.3{\times}$10$^{23}$~cm$^{-2}$ and 
N$_H{\sim}8.7{\times}10^{23}$~cm$^{-2}$, respectively). 
Since Mrk~509\#01 has only an upper limit for the N$_H$ inferred from 
\xray\ analysis ($<1.1{\times}10^{22}$~cm$^{-2}$, see Table~\ref{table1}),
the observed discrepancy for this object might be explained if the source is 
Compton-thick (N$_H{\ge}10^{24}$~cm$^{-2}$) and the observed \xray\ 
spectrum is due to a reflection component. However, we cannot draw any 
firm conclusion on this issue. 

\begin{figure}
\centering
\includegraphics[width=9cm]{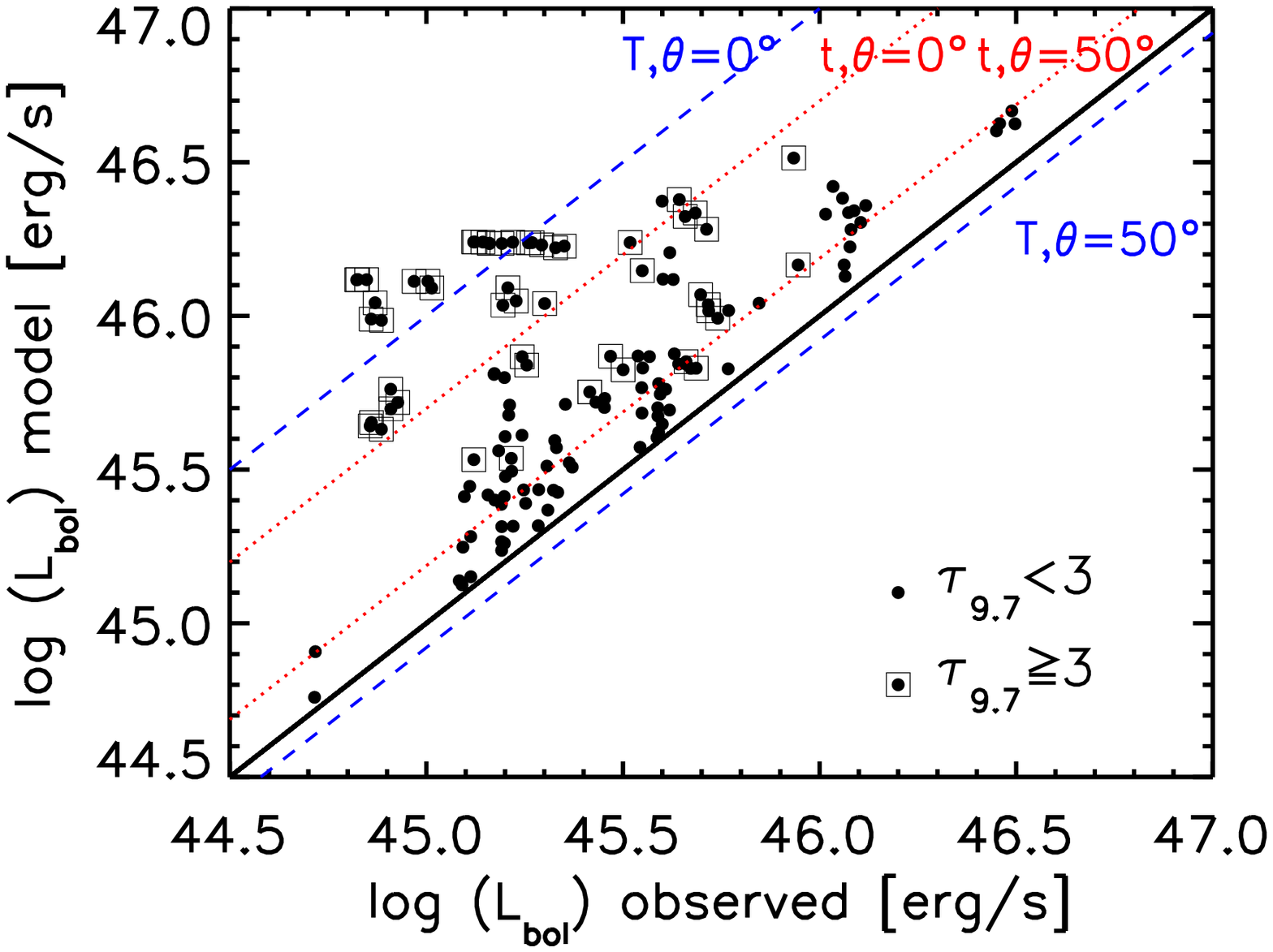}
\caption{`Model' as a function of `observed' bolometric luminosities 
for the sample of luminous obscured quasars at the 1$\sigma$ significance 
level (137 solutions). Squares around the filled circles represent the 
solutions with high optical depth ($\tau_{9.7{\mu}m}\geq{3}$).  The `observed'
bolometric luminosities are reported with no 
corrections applied. The solid line represents the identity relation. 
The dotted and dashed lines represent the predictions from the Pier \& Krolik (1992) model for 4 different configurations for Type~2 
sources as a function of the optical depths $\tau_z$ and $\tau_r$ 
(where $\tau_z$ and $\tau_r$ are the optical depths at 9.7$\mu$m along the 
vertical- with respect the torus equatorial plane - and the radial direction, 
respectively) and viewing angle $\theta$ (with respect to the equatorial 
plane). Red dotted lines: thinner model ($\tau_z=\tau_r=0.1$, labeled with 
the letter t) with $\theta=0,50^\circ$. Blue dashed lines: thicker model 
($\tau_z$=1, $\tau_r$=10, labeled with the letter T) with 
$\theta=0,50^\circ$. At constant optical depth, the configurations with 
smaller viewing angle (closer to the equatorial torus plane) predict lower 
`observed' luminosities.}  

\label{figure_lum}
\end{figure}

\subsection{Host galaxy parameters}
\label{galaxy_fitting}

In the spectral procedure, the host galaxy accounts  for the 
optical/near-IR photometric data points, where the AGN contribution, given 
the obscured nature of our sources, is presumably low. 
We use the SSP spectra weighted by a Schmidt-like law of star formation 
(see Sec. \ref{star_comp_sec}). The extinction E(B-V) and 
$\tau_{\rm sf}$ are free parameters. 
Once the best-fitting values for these two parameters are found, the stellar 
mass (obtained by integrating the star formation history over the
galaxy age and subtracting the fraction due
to mass loss during stellar evolution, $\sim 30$\%, from it) is estimated from the SED 
normalization. 
At the end of the SED fitting procedure, stellar masses are well constrained,
for a given pair of E(B-V) and $\tau_{\rm sf}$, with a typical 1$\sigma$ uncertainty for the normalization of $\sim$~20\%. 

We note that all but one of the stellar 
masses derived in this work are within 30\% from the values estimated 
by Pozzi et al. (2007; eight sources in common), where 
the same data were used but different stellar libraries and a simpler approach was adopted (see Sect. 4.1 and 5.2 of Pozzi et al. 2007). 

The stellar masses found are in the range 4${\times}10^{10}$ up to
5${\times}10^{12}$ M$_{\odot}$ with three very massive galaxies ($>10^{12}$
M$_{\odot}$, see
Table~\ref{table3}),  implying that our obscured AGN are 
hosted by massive galaxies at high redshift.  As said in
Sec. \ref{star_comp_sec}, the masses are obtained using 
a \cite{1955VA......1..283S} initial mass function (IMF) with mass in the 
range (0.15-120 M$_{\odot}$). The assumption of a \cite{2003PASP..115..763C} 
IMF (see \citealt{2006ARA&A..44..141R}) would produce a factor of $\sim{1.7}$ 
lower stellar masses.

In Table~\ref{table3}, the best-fitting value for the free host galaxy
parameters ($\tau_{\rm sf}$ and E(B-V))  and the stellar masses are reported for each source.

\section{Black hole physical properties}

\subsection{Black hole accretion luminosities}
\label{bh_luminosity}         

 The accretion-disk luminosity $L_{acc}$ is a direct output 
of the fitting procedure and is obtained by integrating the code input energy
spectrum once the best-fitting torus components and its normalization are
found (see Sect. 4.3). 

The input spectrum is defined in the 10$^{-3}$ to 20 ${\mu}$m regime. 
Although this wavelength range provides the largest contribution to the 
nuclear AGN luminosity, we have also included in the $L_{acc}$ computation 
the hard \xray\ luminosity ($L_{1.25-500keV}$). 
This luminosity is estimated from the de-absorbed, k-corrected 
$L_{2-10keV}$ luminosity, assumping a photon index $\Gamma$=1.9 (typical of 
AGN emission) and an exponential cut-off at 200 keV 
(e.g., \citealt{2007A&A...463...79G}). 
A different choice for the energy cut-off (e.g., at 100~keV) would produce a 
difference by $\approx$~20\% in the total \xray\ luminosity for $\Gamma$=1.9 
sources (see \citealt{2009arXiv0910.5256V}).

We note that dust grains are almost transparent to hard \xray\ photons, 
therefore the output of the code is not affected by the fact that the 
accretion-disk model spectrum does not extend above soft-\xray\ energies.

\begin{figure}
\centering
\includegraphics[width=9cm]{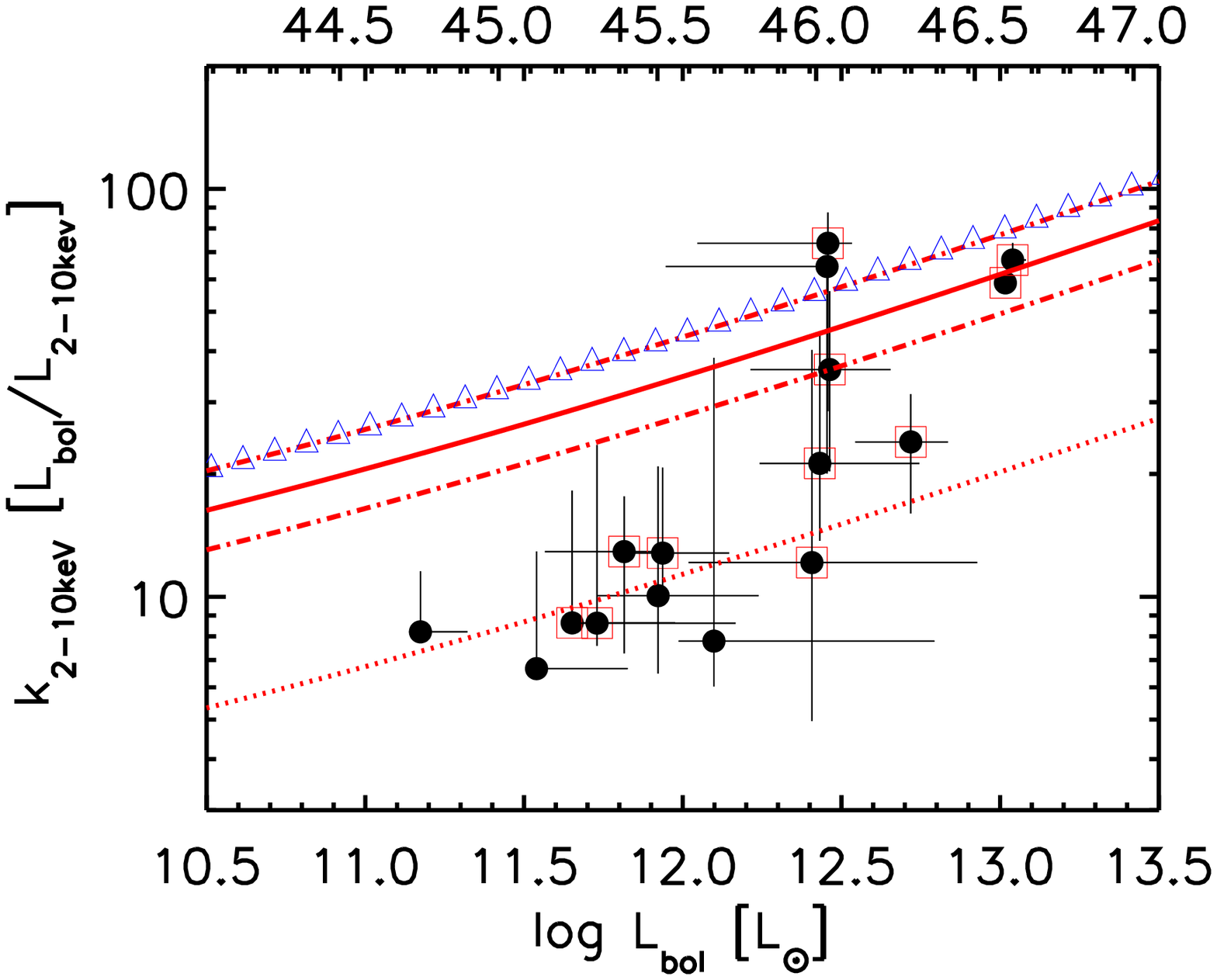}
\caption{2--10 keV bolometric corrections as a function of the `model' 
bolometric luminosities (filled circles). Filled circles inside empty red
squares represent the sources with a spectroscopic redshift. The red solid and dot-dashed 
lines represent the predictions from the 
\cite{2004MNRAS.351..169M} relation and its 1$\sigma$ dispersions. Also the expectations from 
\cite{2007ApJ...654..731H} are reported as empty blue triangles. 
The red dotted line represents the \cite{2004MNRAS.351..169M} 
expectations at 5$\sigma$ from the best-fitting relation.}
\label{figure_kbol}
\end{figure}

In Table~\ref{table3} $L_{bol}$, along with $L_{acc}$ and
$L_{1.25-500keV}$, are reported. $L_{bol}$ extends over two orders of
magnitudes (10$^{44}$-10$^{46}$~erg~s$^{-1}$), with the hard-\xray\
luminosities (1.25-500 keV) contributing in the range 5-50$\%$ of the
AGN power (see Table~\ref{table3}). The two
sources with the highest optical depths ($\tau_{9.7{\mu}m}$=6,10) are among 
the sources with the smallest hard-\xray\ fraction 
(Mrk~509\#01 and PKS~0537\#111). 
In Table~\ref{table3}  we report also the range of bolometric 
luminosities as obtained by considering the full set of $1\sigma$ solutions. 
The uncertainties are, on average, of the order of 0.2 dex, but vary 
significantly from source to source, ranging from about 5\% to about a factor
3 (see also Fig.~\ref{figure_kbol}).


We compare the computed bolometric luminosities with the
luminosities derived by integrating the torus best-fitting templates 
from 0.1-1000$\mu$m (plus adding the hard \xray\
luminosity for self-consistency). The two methods assume the same
torus SED, hence the comparison can give important
information on the systematics affecting the estimates of $L_{bol}$ 
derived by integrating the observed SED, which is the method widely 
used in literature. We refer to the first measures as the `model' 
luminosities and to the latter as the `observed' luminosities. 

The `observed' $L_{bol}$ (see Fig.~\ref{figure_lum}) are lower (up to an 
order of magnitude) than the `model' ones for all the solutions; 
the median value of the ratio is $R{\sim}2$ (the solid line in 
Fig.~\ref{figure_lum} represents the identity relation). 
An under-estimate of the luminosity in Type~2 sources is expected by 
torus models (e.g., \citealt{1993ApJ...418..673P}; 
\citealt{1994MNRAS.268..235G}); here, we quantify this effect and provide 
an empirical factor to correct the `observed' luminosities, at least for 
this class of sources.

We underline how the observed discrepancy does not depend on the lack of
observations at far-IR wavelengths. 
In fact, the two methods assume the same torus SED 
for self-consistency (i.e., the integrated torus SED to estimate the observed
IR luminosity is the output of the code); under this hypothesis, 
an over(under)-estimate on one luminosity would introduce the same effect on the
other. As a result, a poor sampling in the far-IR would have the same impact on both (i.e. `observed'  and  `model' ) luminosities. 
Our analysis takes into account this uncertainty by considering all the 
solutions (i.e., all torus models) at the 1$\sigma$ confidence level. 
By means of this procedure, a broad range of model SED is associated to each 
source (on average, eight solutions; see Sec. \ref{results_fitting}), 
characterized by different emission in the mid/far-IR region, as a result 
of different torus geometry and absorption properties 
(see Fig.~\ref{figure_sed}).

As explained in Pier \& Krolik (1992), the low values of the 
`observed' $L_{bol}$ depend on a combination of three factors: the torus 
opening angle $\Theta$ (geometrical factor), the observer viewing 
angle $\theta$ and the torus optical thickness $\tau_{9.7{\mu}m}$. 
By erroneously assuming isotropic torus emission (as done to compute the 
`observed' $L_{bol}$), the primary flux which does not intercept the 
obscuring material would not be included in the luminosity budget; moreover, as the 
thickness of the torus increases, more and more primary high-energy photons 
entering the torus are absorbed by the dust grains and re-emitted 
isotropically (hence also in directions escaping the torus it self). 
 This effect is explained by the {\it dust self-absorption}, 
i.e. thermal dust emission absorbed by the dust itself. For high 
optical depth, the outer edges of the torus absorb the IR photons coming 
from the warmer dust at smaller radii and re-emit them isotropically, 
i.e. also in directions outside the line of sight.
To better visualize this effect, we report in
Fig.~\ref{figure_lum} the
`observed' versus `model' luminosities, as predicted by Pier \& Krolik (1992), as a function of
the viewing
angle $\theta$ and the
torus optical thickness $\tau_{9.7{\mu}m}$ for 4 sets of Type~2
configurations (as described in Fig.~\ref{figure_lum} caption). 
Although there are some slight differences between the F06 model 
(adopted here) and the Pier \& Krolik (1992) torus model 
(where the optical depth varies independently along the 
radial and the vertical axis), optically thinner models show less
anisotropy (red dotted lines in Fig.~\ref{figure_lum}, corresponding to 
two different viewing angles), than higher $\tau_{9.7{\mu}m}$ models 
(blue dashed lines in Fig.~\ref{figure_lum}, corresponding to the same 
viewing angles considered for the thinner model). The cold outer edges of 
the thicker models, in fact, radiate little and block the light coming from 
the inner torus radii.

To investigate these issues further, 
we apply a `conservative' correction to our `observed' 
luminosities, meant to correct only for the geometrical factor; 
in other words, we divided each `observed' luminosity by the corresponding 
covering factor CF (${\sim}0.58$ for ${\sim}100^{\circ}$ and ${\sim}0.88$ 
for ${\sim}140^{\circ}$). 
Although this correction increases the `observed' luminosities, 
the `model' ones are still higher ($R{\sim}1.6$); the remaining discrepancy 
is mostly found for solutions with high optical depth, 
as expected ($R\sim5$ for models with $\tau_{9.7}{\ge}3$; see 
Fig.~\ref{figure_lum}, where the squares mark the 52 solutions with
$\tau_{9.7}{\ge}3$). 

An independent and consistent analysis was done also by \citet[see their 
$\S$ 5.1]{2007A&A...468..603P} where a first-order correction of ${\sim}2$ 
to the `observed' luminosities was estimated, accounting for geometrical and 
anisotropy effects; in that work, however, the correction was estimated 
using the ratio of obscured/unobscured quasars according to the \cite{2007A&A...463...79G} AGN synthesis 
models of the \xray\ background 
and the different shape of Type~2 vs. Type~1 quasar 
SEDs as a function of the column density. In \cite{2007A&A...468..603P},
the SED fitting was done using the \cite{2004MNRAS.355..973S} AGN templates. Since the template choice was based on
the \xray\ N$_H$ (and not on the N$_H$ resulting from the torus modelling as 
in the present analysis), the correction corresponding to the thicker models 
(N$_H{\gsimeq}10^{24}$~cm$^{-2}$)
were not included since no Compton-thick objects were revealed in X-rays.

\subsection{Hard \xray\ bolometric corrections}
\label{bolometric_correction}
       
In Fig.~\ref{figure_kbol} the bolometric-to-\xray\ luminosity ratio
($k_{2-10keV}$) as a function of
$L_{bol}$ is shown: for the bolometric luminosities we assume the
model ones. The error bars on $k_{2-10keV}$ are derived from the 1$\sigma$ 
dispersion on $L_{bol}$.

A large spread in the $k_{2-10keV}$ is found
($6{\lsimeq}k_{2-10keV}{\lsimeq}80$), as pointed out also by the
pioneering work of \cite{1994ApJS...95....1E} on Type~1 QSOs, due to the
large dispersion in the AGN spectral shape. Our median value
($k_{2-10keV}\sim{20}$, estimated on the 137 solutions),
is marginally consistent with the mean value of
\cite{1994ApJS...95....1E}, ${\sim}25$ after removing the IR
contribution (in order to not double-count the fraction of the
nuclear emission absorbed by the circumnuclear dusty material seen
almost face-on). We confirm the trend of higher $k_{2-10keV}$ for objects with
higher bolometric luminosities as predicted by
\cite{2004MNRAS.351..169M} (red solid line in
Fig.~\ref{figure_kbol}; red dot-dashed lines representing the 1$\sigma$
model dispersion), but our $k_{2-10keV}$ values are
significantly lower (at least a factor 2 in normalization). They derive $k_{2-10keV}$ by 
constructing an AGN reference template taking into account how the
  spectral index $\alpha_{ox}$ varies as a function of the 
luminosity (\citealt{2003AJ....125..433V}). Predictions consistent with 
\cite{2004MNRAS.351..169M} were obtained more recently
by Hopkins et al. (2007, blue triangles in Fig.~\ref{figure_kbol}), 
considering the most recent
determination of SED templates (i.e. \citealt{2006ApJS..166..470R}) and
$\alpha_{ox}$ (i.e. \citealt{2006AJ....131.2826S}). 

Our low values for $k_{2-10keV}$ are consistent with our previous analysis
(median $k_{2-10keV}\sim{25}$, \citealt{2007A&A...468..603P}) based on 
a different method and on different AGN templates (\citealt{2004MNRAS.355..973S}) and with other 
estimates found in literature for hard \xray\
selected samples. \cite{2003ApJ...590..128K}, considering a sample
of \xray\ selected luminous AGN 
($10^{43}<L_{2-10keV}<10^{46}$~erg~sec$^{-1}$) found a median $k_{2-10keV}$ 
of 18; \cite{2007ApJ...667...97B}, analysing a sample of low-luminosity AGN
($10^{42}<L_{2-10keV}<10^{43.6}$~erg~sec$^{-1}$), found a median 
$k_{2-10keV}$ of 12.
Low bolometric-to-\xray\ ratios, consistent with our estimate
(median $k_{2-10keV}\sim{25}$,  1$\sigma$=53) were found recently by \cite{2009arXiv0912.4166L}, 
where the statistical properties of a large (and complete) sample of 545 \xray\ selected Type~1 QSO from
the XMM-COSMOS survey (\citealt{2007ApJS..172...29H}) are presented.

The lower bolometric-to-\xray\ luminosity ratios found in the present
work (and in the above mentioned samples), in
comparison to the \cite{2004MNRAS.351..169M} and \cite{2007ApJ...654..731H} 
predictions, are probably caused by a 
selection bias, since our sample (and most of the above cited ones) are 
hard-\xray\ selected samples (i.e., sources with high \xray\ luminosity are 
favored). Moreover, as discussed in $\S$ \ref{sample_sec}, our sources are 
among the most extreme \xray\ sources,
being characterized by red optical-to-NIR colours (R-K$_{s}{\gsimeq}5$) and
high \xray  -to-optical ratio (X/O$\gsimeq{1}$). Our selection
is likely the origin of the large deviation (at about the 5$\sigma$
level) for a large fraction of the present sample (see Fig.~\ref{figure_kbol})
from the \cite{2004MNRAS.351..169M} relation. To further explore this issue, 
a larger (and complete) sample of \xray\ sources (with optical identification
up to the faintest \xray\ fluxes) is needed, in order to correct for the 
selection bias and to derive the properties of the parent AGN population 
(see \citealt{2009arXiv0912.4166L}).

\subsection{Black hole masses}

\begin{figure}
\centering
\includegraphics[width=10cm]{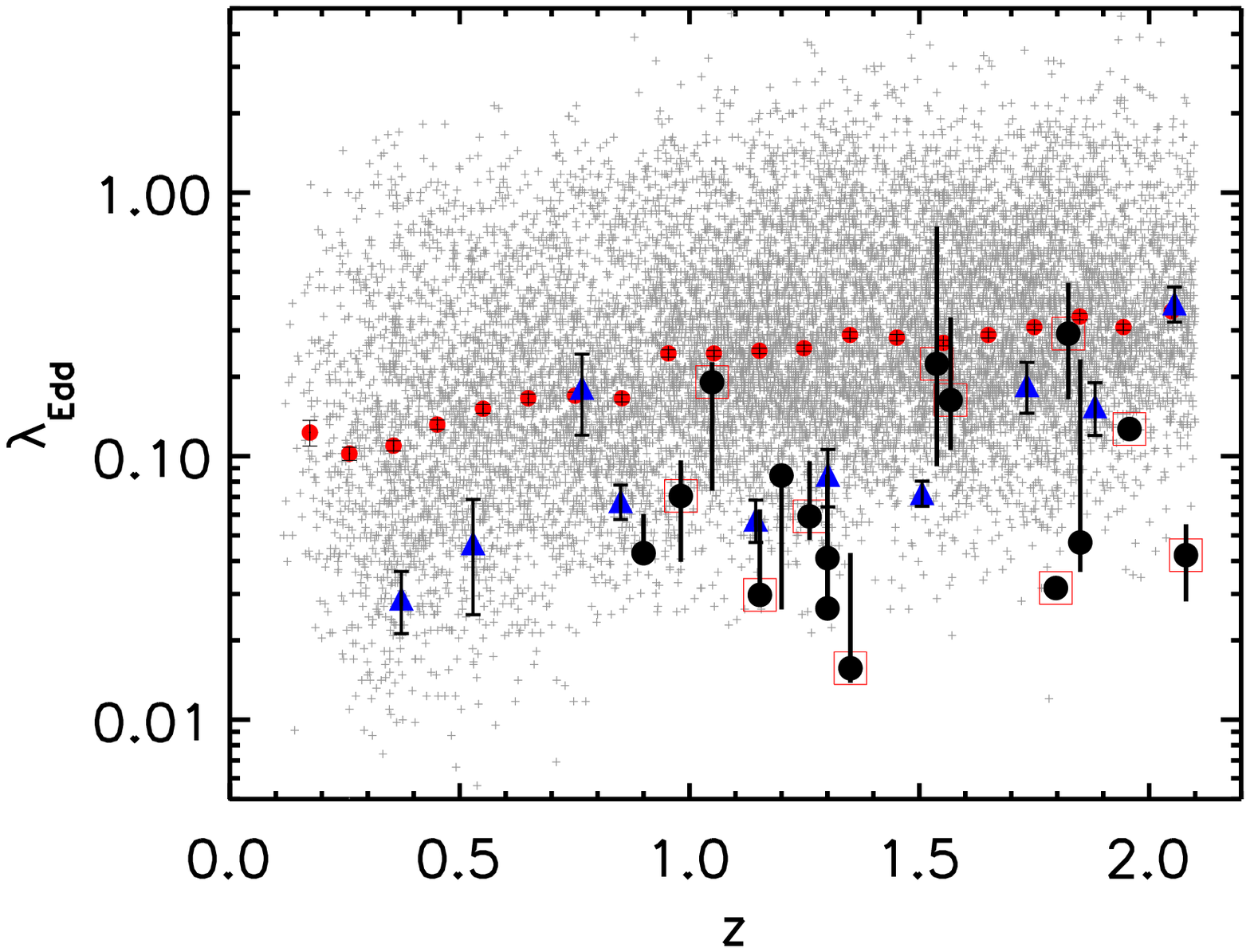}
\caption{$\lambda_{Edd}$ as a function of $z$. Black circles: 
sources of the present sample. Red squares as in Fig.7.  The error bars represent the 1$\sigma$ uncertainties 
on L$_{bol}$ (as derived from the $\chi^{2}$ analysis). Small grey crosses: 
sample of SDSS quasars from \cite{2004MNRAS.352.1390M}, with the median values
(and associated uncertainties assuming the normalized median absolute values)  of
  $\lambda_{Edd}$ within ${\Delta}z=0.1$ bins shown as red filled
  circles.  Blue triangles: median values (and associated uncertainties assuming the normalized median absolute values) of
  $\lambda_{Edd}$ within ${\Delta}z=0.2$ bins from the COSMOS Type 1 sample of \cite{2009arXiv0912.4166L}.}
\label{figure_mclure}
\end{figure}

The BH masses are not a direct output of the best-fitting
procedure and cannot be derived using `standard methods' (i.e. galaxy
stellar kinematics, nuclear gas motions, reverberation).
We estimate them indirectly using the $M_{bulge}-M_{BH}$ relation
derived locally by \cite{2003ApJ...589L..21M}, by assuming as
$M_{bulge}$ the stellar mass derived from our best-fitting procedure. 
The main uncertainties affecting these estimates derive from the extrapolation 
of the local relation to higher $z$, where the behaviour of this relation is
still a matter of debate (see discussion in \citealt{2007A&A...468..603P}). 
As far as the stellar masses are concerned, they are quite well constrained 
by the SED-fitting procedure inside the pre-constructed grid of galaxy models
(see Sec. \ref{galaxy_fitting}). 

The inferred black hole masses are typically in the range
$10^8-10^9$M$_{\odot}$, with three sources (PKS~0537$\#$43,
GD~158$\#$19, Abell~2690$\#$29) with higher masses
(M$_{BH}{\sim}10^{9.5}-10^{10.0}$M$_{\odot}$). 

The range of BH masses is consistent with the values 
reported by \cite{2004MNRAS.352.1390M} for the SDSS quasars in the same 
redshift interval ($0.9{\lsimeq}z{\lsimeq}2.1$, see also 
\citealt{2008ApJ...680..169S}, where new BH masses are derived). 

In Fig.~\ref{figure_mclure}, the Eddington ratios $\lambda_{Edd}$,
defined as $\lambda_{Edd}=L_{bol}/L_{Edd}$ (with
$L_{Edd}=1.38{\times}10^{38}M_{BH}/M_{\odot}$), are reported as a
function of redshift. The values are compared
with those of the whole SDSS quasar sample 
(small grey crosses, \citealt{2004MNRAS.352.1390M}) and those obtained  by
 \cite{2009arXiv0912.4166L} for the sub-sample of 150  X-ray selected Type 1 AGN in
COSMOS with an accurate black hole mass determination (blue triangles).

The $\lambda_{Edd}$ values of the present work cover slightly more
than an order of magnitude (0.01-0.3), with a median value of
$\lambda_{Edd}{\sim}0.08$ (estimated considering all the 137 solutions at
1$\sigma$ level, see Sec. \ref{results_fitting}). The derived values are within the 3$\sigma$ confidence interval of the SDSS 
quasar $\lambda_{Edd}$ distribution, characterized by a median value 
of ${\sim}0.3$ and with a dispersion of 0.35 dex at the same redshift interval
sampled by our sources. However, almost 
all our data points lie
towards the low $\lambda_{Edd}$ tail of the SDSS distribution (see
Fig.~\ref{figure_mclure}), suggesting
that \xray\ selection is biased towards slightly lower $\lambda_{Edd}$
than optical selection. Our data are fully consistent with the results
  obtained from a much larger sample of X-ray
selected Type 1 AGN in the COSMOS field  (\citealt{2009arXiv0912.4166L}).

The results are robust
against the uncertainties on the extrapolation, discussed above, of the
local $M_{bulge}-M_{BH}$ relation, at the redshift of our
sample. In fact, allowing for positive evolution with
redshift of the $M_{BH}/M_{bulge}$
ratio by a factor of 2 (e.g., \citealt{2006ApJS..163....1H}; \citealt{2006NewAR..50..809S}; \citealt{2010ApJ...708..137M}), the Eddington ratios $\lambda_{Edd}$
would decrease further by the same factor. 

\begin{figure}
\centering
\includegraphics[width=10cm]{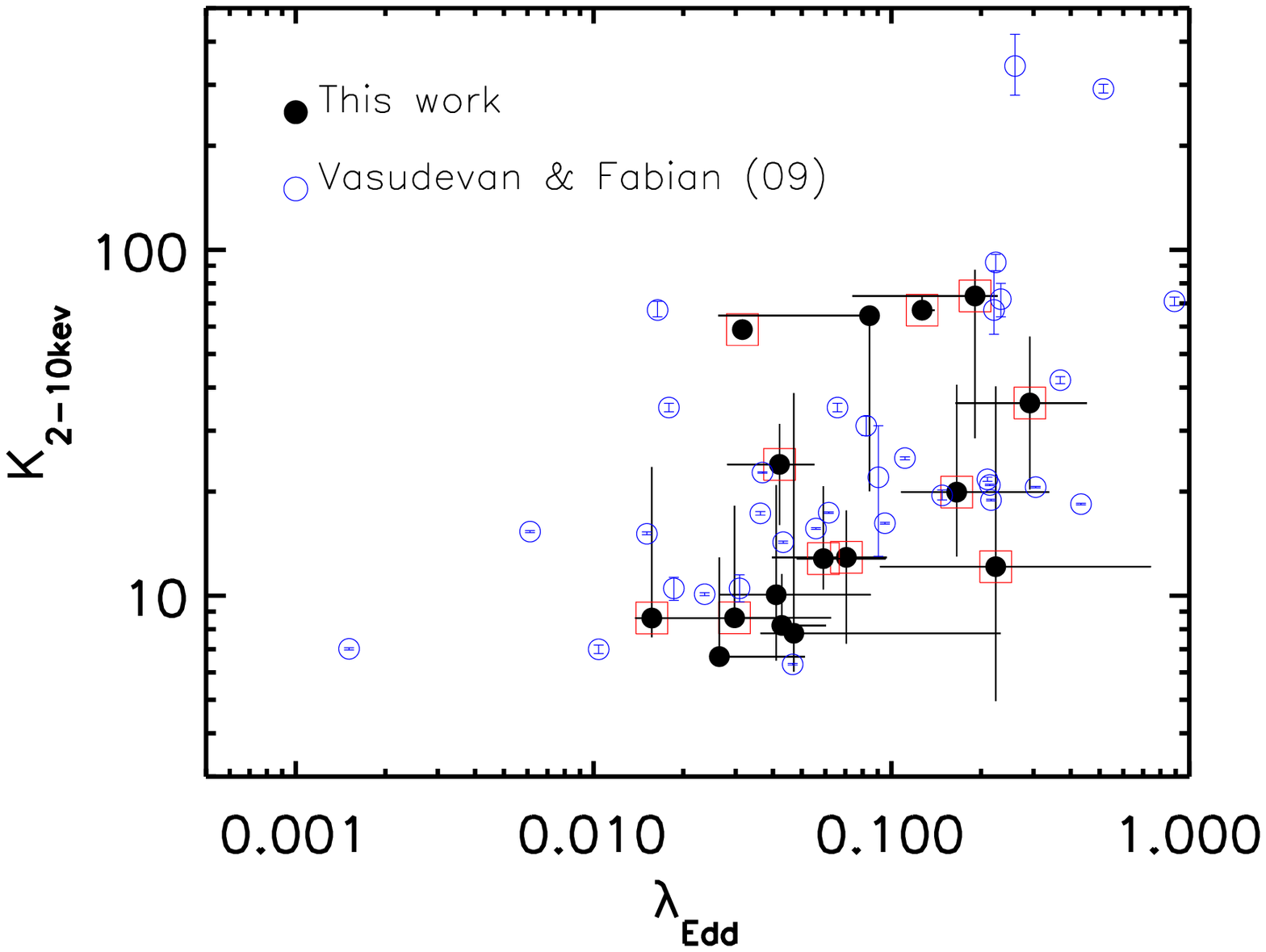}
\caption{$k_{2-10keV}$ as a function of $\lambda_{Edd}$. Black filled symbols: 
sources of the present sample.  Red squares as in Fig.7. The error bars represent the 1$\sigma$ 
uncertainties on L$_{bol}$ (as derived from the $\chi^{2}$ analysis) which 
affect both the $k_{2-10keV}$ and the $\lambda_{Edd}$ values. 
Blue open symbols: sources from \cite{2009MNRAS.392.1124V}. }
\label{figure_vasudevan}
\end{figure}

Finally, in Fig.~\ref{figure_vasudevan}, the bolometric corrections 
$k_{2-10keV}$ are plotted against the Eddington ratios $\lambda_{Edd}$ 
(following \citealt{2009MNRAS.392.1124V}). The sources of the present
work are reported as black filled circle (the error bars representing
the $1\sigma$ confidence interval derived from the uncertainties on
$L_{bol}$). Along with our data we show the \cite{2009MNRAS.392.1124V} results, where simultaneous optical, UV and \xray\ 
observations are included for the majority of the \cite{2004ApJ...613..682P} 
reverberation mapped sample of AGN (blue empty circles).
Our findings are in fairly good agreement with the trend of increasing $k_{2-10keV}$ for increasing $\lambda_{Edd}$. 
\cite{2009MNRAS.392.1124V} interpret the observed trend as due to
 different black hole SED shape as a function of
the Eddington ratio, with the high and low Eddington ratios
corresponding to different fractions of the ionizing UV
luminosity. 
 A similar trend was recently found for a sample of 63 
Type~1 and Type~2 AGN detected in the \swift/BAT 9-months catalog (see \citealt{2009arXiv0910.5256V}). At variance with the assumption in \cite{2009MNRAS.392.1124V}, where the bolometric luminosities were 
derived by integration over the observed optical/UV/X-ray SED, 
in this work the authors consider the reprocessed IR emission, reproduced by 
the empirical  SEDs of Silva et al. (2004), as a proxy of the 
intrinsic AGN bolometric luminosity, as firstly suggested by Pozzi et al. (2007). 

The dependence of bolometric corrections on Eddington ratios is expected by
accretion-disk models,  which predict an increasing hard \xray\
bolometric corrections at increasing accretion rates (e.g. \citealt{1997MNRAS.286..848W}).
Recently also \cite{2009A&A...501..915B}, studying a
large (156 sources) sample of Type~1 \xray\ AGN from the XMM-$Newton$
archive suggest that the  bolometric correction must depend on Eddington ratio in order to allow the intrinsic power of
AGN to scale linearly with black hole masses. 

\begin{sidewaystable*}
\caption{Best-fitting physical parameters and inferred rest-frame properties.}
\label{table3}
\centering
\begin{tabular}{l cc ccc ccc ccc cccc  c } \\\hline\hline 
  Source name         & $\chi^{2}_{min}$/$d.o.f$     &  $\alpha$& ${\tau}_{9.7{\mu}m}$ &
  $\Theta$&  $\tau_{sf}$   & $E(B-V)$   & $L_{acc}$ & $L_{1-1000_{{\mu}m}}$ &$L_{2-10keV}$
  &$L_{1.25-500keV}$ & $L_{bol}$ & $k_{2-10keV}$ &  $M_{star}$ & $M_{BH}$ & $\lambda_{Edd}$   \\

(1)          &            (2)       &       (3)        &  (4) &    (5)    &
(6)   & (7) &(8) & (9)   & (10) &(11)  & (12) &(13) & (14) & (15)& (16) \\\hline    

PKS~0537\#43 &  66.7/6 & -1.0 &   0.1 &    140  & 0.05 & 0.3 & 373.6 &    254 &6.8
&  26.3 &  400 [400-422] &  58.8& 49.4 & 10.05 & 0.032   \\

PKS~0537\#11 & 7.8/5 &  0.0   &   2.0   &    140 & 0.7  & 0.64 &  17.6 &      7.4 &
 1.9 &  7.6     & 25.2 [14.1-34.4] &     12.9 & 1.20 & 0.28 & 0.070 \\

PKS~0537\#164 &      9.0/5 &     0.0 &      3.0 &     60 & 0.15 &0.0 &    
99.8 &      4.9 &      3.1 &     12.1 & 112 [63.0-174] &     36.1 &1.29 & 0.30& 0.292 \\

PKS~0537\#123 &      5.0/5 &           0.0 &      0.1 &    140 &1.0 &0.65 &
9.5 &      7.8 &      2.0 &      7.8  & 17.3 [17.3-36.4] &      8.6 &1.99 & 0.46 & 0.030  \\

GD~158\#62 &      9.3/5 &   0.0 & 0.1 &    100 & 0.35 &0.44 & 85.1 &    39.7 &
4.9 &     19.0 & 104 [67.3-214] &     21.2 & 2.19 & 0.51 & 0.163 \\

GD~158\#19 &      26.9/6 &           0.0 &      0.3 &    140 & 1.0&0.49 &
397. &      289.7&      6.3 &     24.4&
422 [422-465] &     67.0 & 12.3 & 2.64 & 0.127 \\

Mrk~509\#01    &     17.9/5&        -0.5 &      6.0 & 60 & 0.9    &0.85 &
104.5 &      1.59 &      1.5 &      5.8&    111 [42.8-131] &
73.6 &1.98 & 0.46& 0.191 \\

Mrk~509\#13   &     25.0/5 &           0.0 &      1.0 &    140   &0.6 & 1.0 &     23.2 &      13.0 &      2.6 &      10.1 &
$33.3 [27.1-53.9]$  &     12.8 &  1.93 & 0.45& 0.059 \\

Abell~2690\#75 &      17.5/5 &           0.0 &      1.0 & 140 & 0.05 & 0.18&     19.8&      11.0 &      3.2 &
12.4 &     32.2 [20.8-66.8] &     10.1 &2.73 & 0.62& 0.041 \\

PKS~0312\#36 &      9.6/5  &           0.0 &      0.1 &    140 & 0.85& 0.79 &
3.0 &      2.5 &      0.7 &      2.7 & 5.7 [5.7-8.1] &     8.2&  0.43 & 0.11& 0.043  \\

PKS~0537\#91 &      2.9/5 &           0.0 &      3.0 &    140 &0.25 &0.3 &       66.0 &      23.7 &      8.1 &     31.4 &
98.3 [40.2-326] &      12.1 &  1.48 &0.35& 0.224 \\

PKS~0537\#54 &    2.9/5 &           0.0 &      0.1 &    140 &0.05 &0.17 &       5.6 &      4.6 &      2.0 &      7.8 &
13.3 [13.3-25.8] &      6.7 &  1.72 & 0.40& 0.026 \\

PKS~0537\#111  &      11.4/5 &           -0.5 &     10.0 &  140 & 0.05 &0.31 &      103.1 &      13.4 &      1.7 &
6.6 &    110 [34.1-110] &     64.4 &  4.61 & 1.03& 0.084 \\

Abell~2690\#29   &      0.5/2 &           0.0 &      1.0& 140 &  0.15& 0.55&   168.8 &     94.7 &      8.4 &
32.6 &    201 [135-264]  &     23.9 & 17.9 & 3.80 &  0.042  \\

PKS~0312\#45  &      0.5/5 &           0.0 &      0.1 &    100 & 0.25  & 0.48& 24.25&     11.3 &      6.2 &   24.0 &
48.3 [37.4-239] &      7.8 & 3.61 & 0.81&  0.047 \\

BPM~16274\#69 &      0.5/5 &           0.0 &      0.6 &    140 & 0.05 & 0.25  &  11.4 &      7.3 &      2.4 &      9.3 &
20.7 [18.2-56.6] &      8.6 & 4.68 & 1.05& 0.016 \\\hline
\end{tabular}

\begin{center}
\begin{minipage}[h]{22cm}
\footnotesize
(1) source name\\
 (2) best-fitting minimum $\chi^{2}$ ($\chi^{2}_{min}$) and number of parameters to be fitted (degrees of freedom); \\
(3), (4), (5) best-fitting torus parameters ($\alpha$: exponent of the power law parameterizing the density profile; $\tau_{9.7{\mu}m}$: 9.7 ${\mu}$m optical depth, $\Theta$: torus
opening angle). The ratio parameter $R_{max}/R_{min}$ is frozen to 30; the
density parameter $\gamma$ is frozen to 0; \\
 (6), (7) best-fitting stellar parameters ($\tau_{sf}$: duration of the
exponential decay of the burst in units of the oldest SSP; $E(B-V)$: extinction);  \\
(8) accretion-disk model luminosity, $L_{acc}$ (from soft \xray\ to
IR frequencies) which represents the torus model input luminosity, in units of
10$^{44}$ erg s$^{-1}$;\\
(9) integrated (1-1000$\mu$m) torus
luminosity in units of
10$^{44}$ erg s$^{-1}$ (not corrected, see \ref{bh_luminosity}); \\
(10) absorption-corrected 2--10 keV luminosity in units of
10$^{44}$ erg $s^{-1}$; \\
(11) hard-\xray\ (1.25-500 keV) luminosity in units of
10$^{44}$ erg s$^{-1}$ (derived from the 2--10 keV
luminosity, see \ref{bolometric_correction}); \\
(12) bolometric AGN luminosity  ($L_{acc}$+ $L_{1.25-500keV}$) in units of
10$^{44}$ erg s$^{-1}$; the 1$\sigma$ range derived from the SED-fitting
analysis is reported;  \\
 (13) 2--10 keV bolometric correction ($L_{bol}/L_{2-10keV}$); \\
(14) galaxy mass in units of 10$^{11}$M$_{\odot}$; \\
(15) black hole masses in units of 10$^{9}$M$_{\odot}$ (estimated from \citealt{2003ApJ...589L..21M}
relation); \\
(16) Eddington ratios
($L_{bol}/L_{Edd}$).
\end{minipage}
\end{center}
\end{sidewaystable*}

\section{Summary}

We analyzed the SEDs of a sample of 16 obscured quasars selected in the 
hard \xray\ band. \spitzer\ mid/far-IR photometry (IRAC and MIPS), along 
with the data available in the literature, is modeled using a multi-component model, where the AGN re-processed emission is reproduced in the context of 
a flared disk model, as described by F06. 
Within the context of a flared disk torus model, the uncertainty in and
degeneracy between the various derived parameters are accounted for by
including all solutions within 1$\sigma$ of the best- fit in the subsequent
analysis.

The main results are summarized below: 

\begin{itemize}

\item[$\bullet$]
{All the 16 quasars are detected up to 8 ${\mu}$m and 
all, but two sources, are detected at 24${\mu}$m with flux densities 
in the range 100-7000 ${\mu}$Jy at the 5$\sigma$ level. 
The two most luminous sources of the sample are detected also at 70 
and 160 ${\mu}$m.}

\item[$\bullet$]
{The observed broad-band spectral energy distributions are 
well reproduced by a multi-component model comprising a stellar, an AGN and a 
starburst components (when far-IR detections are available). 
The AGN component, modelled with the F06 radiative transfer code, accounts 
for the \xray\ emission and for a fraction of the IR emission, mainly due to 
reprocessed emission from the putative dusty torus surrounding the central 
black hole.  }

\item[$\bullet$]
{Solutions with a moderate optical depth $\tau_{9.7}$ are favoured by the
  SED-fitting, with the majority of the 
sources having moderate optical depths (${\tau}_{9.7{\mu}m}{\leq}3$). 
The derived gas column densities (N$_{H}$) are consistent, for most of 
the sources, with the values estimated from the \xray\ analysis, both 
indicating that the sources are Compton-thin AGN 
(N$_{H}{\sim}10^{22}-3{\times}10^{23}$cm$^{-2}$).}

\item[$\bullet$]
{The model nuclear bolometric luminosities are in the range 
$5{\times}10^{44}-4{\times}10^{46}$~erg~s$^{-1}$. By comparing these  values 
with those obtained by the integration of the nuclear observed SED, we 
conclude that the latter under-estimate the bolometric luminosities 
by a factor of 2. The difference may be explained by anisotropic torus 
emission and the effect of the torus optical depth 
(e.g., Pier \& Krolik 1992).}

\item[$\bullet$]
{From the model nuclear SEDs, we estimate the
bolometric-to-\xray\ corrections ($k_{2-10keV}$). The median $k_{2-10keV}$ is 
${\sim}20$ ($6{\lsimeq}k_{2-10keV}{\lsimeq}80$). The value is smaller than 
assumed by some models of BH evolution ($k_{2-10keV}{\gsimeq}40$ at the 
median luminosity of our sample). 
The discrepancy is significant at 5$\sigma$ level at low bolometric 
luminosity.}

\item[$\bullet$]
{By assuming the local $M_{bulge}-M_{BH}$ relation, 
we estimate $\lambda_{Edd}$ with a median value of 0.08 
($0.01{\lsimeq}\lambda_{Edd}{\lsimeq}0.3$). The whole SDSS quasar sample, at the same redshift interval sampled by our 
objects, is characterized by a median value of 0.3.  Our data are within the  $3\sigma$ confidence level of the optically 
selected quasar distribution. However, almost all our sources lie towards 
the low $\lambda_{Edd}$ tail of the SDSS distribution, suggesting that our 
\xray\ selection is biased towards lower Eddington efficiencies than optical 
selection.}

\item[$\bullet$]
{The data are consistent with the correlation recently suggested 
by Vasudevan \& Fabian (2007, 2009) between $k_{2-10keV}$ and 
$\lambda_{Edd}$, where low bolometric corrections are found at low Eddington 
ratios.}

\end{itemize}

\begin{acknowledgements}
The authors thank the anonymous referee for useful comments that helped improve both the paper content and presentation.
 The authors thank D. Fadda for suggestions about MIPS data reduction 
techniques, R. Gilli and F. La Franca for helpful 
discussions and E. Lusso and E. Sarria for providing their results 
before publication. 
This work has benefited for partial support by the Italian Space Agency
(contracts ASI/COFIS/WP3110 I/026/07/0 and ASI I/088/06/0), 
PRIN/MIUR (grant 2006-02-5203) and from research
funding from the European Community's Sixth Framework Programme under RadioNet contract R113CT 2003 5158187.
This work is based on observations  made with the \spitzer\ 
Space Telescope, which is operated by the Jet Propulsion Laboratory,
California Institute of Technology under NASA contract 1407 and with
the \scuba\ camera, operating at the James Clerk
Maxwell Telescope, a joint U.K.-Dutch-Canadian
millimeter/sub-millimeter Telescope in Hawaii.

\end{acknowledgements}

\bibliographystyle{aa}

\bibliography{pozzi}

\begin{thebibliography}{73}
\expandafter\ifx\csname natexlab\endcsname\relax\def\natexlab#1{#1}\fi

\bibitem[{{Agol} {et~al.}(2009){Agol}, {Gogarten}, {Gorjian}, \&
  {Kimball}}]{2009ApJ...697.1010A}
{Agol}, E., {Gogarten}, S.~M., {Gorjian}, V., \& {Kimball}, A. 2009, \apj, 697,
  1010

\bibitem[{{Baldi} {et~al.}(2002){Baldi}, {Molendi}, {Comastri}, {Fiore},
  {Matt}, \& {Vignali}}]{2002ApJ...564..190B}
{Baldi}, A., {Molendi}, S., {Comastri}, A., {et~al.} 2002, \apj, 564, 190

\bibitem[{{Ballo} {et~al.}(2007){Ballo}, {Cristiani}, {Fasano}, {Fontanot},
  {Monaco}, {Nonino}, {Pignatelli}, {Tozzi}, {Vanzella}, {Fontana},
  {Giallongo}, {Grazian}, \& {Danese}}]{2007ApJ...667...97B}
{Ballo}, L., {Cristiani}, S., {Fasano}, G., {et~al.} 2007, \apj, 667, 97

\bibitem[{{Berta} {et~al.}(2004){Berta}, {Fritz}, {Franceschini}, {Bressan}, \&
  {Lonsdale}}]{2004A&A...418..913B}
{Berta}, S., {Fritz}, J., {Franceschini}, A., {Bressan}, A., \& {Lonsdale}, C.
  2004, \aap, 418, 913

\bibitem[{{Bianchi} {et~al.}(2009){Bianchi}, {Bonilla}, {Guainazzi}, {Matt}, \&
  {Ponti}}]{2009A&A...501..915B}
{Bianchi}, S., {Bonilla}, N.~F., {Guainazzi}, M., {Matt}, G., \& {Ponti}, G.
  2009, \aap, 501, 915

\bibitem[{{Brusa} {et~al.}(2005){Brusa}, {Comastri}, {Daddi}, {Pozzetti},
  {Zamorani}, {Vignali}, {Cimatti}, {Fiore}, {Mignoli}, {Ciliegi}, \&
  {R{\"o}ttgering}}]{2005A&A...432...69B}
{Brusa}, M., {Comastri}, A., {Daddi}, E., {et~al.} 2005, \aap, 432, 69

\bibitem[{{Cardelli} {et~al.}(1989){Cardelli}, {Clayton}, \&
  {Mathis}}]{1989ApJ...345..245C}
{Cardelli}, J.~A., {Clayton}, G.~C., \& {Mathis}, J.~S. 1989, \apj, 345, 245

\bibitem[{{Chabrier}(2003)}]{2003PASP..115..763C}
{Chabrier}, G. 2003, \pasp, 115, 763

\bibitem[{{Cocchia} {et~al.}(2007){Cocchia}, {Fiore}, {Vignali}, {Mignoli},
  {Brusa}, {Comastri}, {Feruglio}, {Baldi}, {Carangelo}, {Ciliegi}, {D'Elia},
  {La Franca}, {Maiolino}, {Matt}, {Molendi}, {Perola}, \&
  {Puccetti}}]{2007A&A...466...31C}
{Cocchia}, F., {Fiore}, F., {Vignali}, C., {et~al.} 2007, \aap, 466, 31

\bibitem[{{Dullemond} \& {van Bemmel}(2005)}]{2005A&A...436...47D}
{Dullemond}, C.~P. \& {van Bemmel}, I.~M. 2005, \aap, 436, 47

\bibitem[{{Efstathiou} \& {Rowan-Robinson}(1995)}]{1995MNRAS.273..649E}
{Efstathiou}, A. \& {Rowan-Robinson}, M. 1995, \mnras, 273, 649

\bibitem[{{Elitzur}(2008)}]{2008NewAR..52..274E}
{Elitzur}, M. 2008, New Astronomy Review, 52, 274

\bibitem[{{Elitzur} \& {Shlosman}(2006)}]{2006ApJ...648L.101E}
{Elitzur}, M. \& {Shlosman}, I. 2006, \apjl, 648, L101

\bibitem[{{Elvis} {et~al.}(1994){Elvis}, {Wilkes}, {McDowell}, {Green},
  {Bechtold}, {Willner}, {Oey}, {Polomski}, \& {Cutri}}]{1994ApJS...95....1E}
{Elvis}, M., {Wilkes}, B.~J., {McDowell}, J.~C., {et~al.} 1994, \apjs, 95, 1

\bibitem[{{Fabian}(1999)}]{1999MNRAS.308L..39F}
{Fabian}, A.~C. 1999, \mnras, 308, L39

\bibitem[{{Fadda} {et~al.}(2006){Fadda}, {Marleau}, {Storrie-Lombardi},
  {Makovoz}, {Frayer}, {Appleton}, {Armus}, {Chapman}, {Choi}, {Fang},
  {Heinrichsen}, {Helou}, {Im}, {Lacy}, {Shupe}, {Soifer}, {Squires}, {Surace},
  {Teplitz}, {Wilson}, \& {Yan}}]{2006AJ....131.2859F}
{Fadda}, D., {Marleau}, F.~R., {Storrie-Lombardi}, L.~J., {et~al.} 2006, \aj,
  131, 2859

\bibitem[{{Fiore} {et~al.}(2003){Fiore}, {Brusa}, {Cocchia}, {Baldi},
  {Carangelo}, {Ciliegi}, {Comastri}, {La Franca}, {Maiolino}, {Matt},
  {Molendi}, {Mignoli}, {Perola}, {Severgnini}, \&
  {Vignali}}]{2003A&A...409...79F}
{Fiore}, F., {Brusa}, M., {Cocchia}, F., {et~al.} 2003, \aap, 409, 79

\bibitem[{{Fiore} {et~al.}(2008){Fiore}, {Grazian}, {Santini}, {Puccetti},
  {Brusa}, {Feruglio}, {Fontana}, {Giallongo}, {Comastri}, {Gruppioni},
  {Pozzi}, {Zamorani}, \& {Vignali}}]{2008ApJ...672...94F}
{Fiore}, F., {Grazian}, A., {Santini}, P., {et~al.} 2008, \apj, 672, 94

\bibitem[{{Frayer} {et~al.}(2006){Frayer}, {Huynh}, {Chary}, {Dickinson},
  {Elbaz}, {Fadda}, {Surace}, {Teplitz}, {Yan}, \&
  {Mobasher}}]{2006ApJ...647L...9F}
{Frayer}, D.~T., {Huynh}, M.~T., {Chary}, R., {et~al.} 2006, \apjl, 647, L9

\bibitem[{{Frayer} {et~al.}(2009){Frayer}, {Sanders}, {Surace}, {Aussel},
  {Salvato}, {Le Floc'h}, {Huynh}, {Scoville}, {Afonso-Luis}, {Bhattacharya},
  {Capak}, {Fadda}, {Fu}, {Helou}, {Ilbert}, {Kartaltepe}, {Koekemoer}, {Lee},
  {Murphy}, {Sargent}, {Schinnerer}, {Sheth}, {Shopbell}, {Shupe}, \&
  {Yan}}]{2009arXiv0902.3273F}
{Frayer}, D.~T., {Sanders}, D.~B., {Surace}, J.~A., {et~al.} 2009, ArXiv
  e-prints

\bibitem[{{Fritz} {et~al.}(2006){Fritz}, {Franceschini}, \&
  {Hatziminaoglou}}]{2006MNRAS.366..767F}
{Fritz}, J., {Franceschini}, A., \& {Hatziminaoglou}, E. 2006, \mnras, 366, 767
  (F06)

\bibitem[{{Gilli} {et~al.}(2007){Gilli}, {Comastri}, \&
  {Hasinger}}]{2007A&A...463...79G}
{Gilli}, R., {Comastri}, A., \& {Hasinger}, G. 2007, \aap, 463, 79

\bibitem[{{Granato} \& {Danese}(1994)}]{1994MNRAS.268..235G}
{Granato}, G.~L. \& {Danese}, L. 1994, \mnras, 268, 235

\bibitem[{{Gruppioni} {et~al.}(2008){Gruppioni}, {Pozzi}, {Polletta},
  {Zamorani}, {La Franca}, {Sacchi}, {Comastri}, {Pozzetti}, {Vignali},
  {Lonsdale}, {Rowan-Robinson}, {Surace}, {Shupe}, {Fang}, {Matute}, \&
  {Berta}}]{2008ApJ...684..136G}
{Gruppioni}, C., {Pozzi}, F., {Polletta}, M., {et~al.} 2008, \apj, 684, 136

\bibitem[{{Haardt} \& {Maraschi}(1991)}]{1991ApJ...380L..51H}
{Haardt}, F. \& {Maraschi}, L. 1991, \apjl, 380, L51

\bibitem[{{Hasinger} {et~al.}(2007){Hasinger}, {Cappelluti}, {Brunner},
  {Brusa}, {Comastri}, {Elvis}, {Finoguenov}, {Fiore}, {Franceschini}, {Gilli},
  {Griffiths}, {Lehmann}, {Mainieri}, {Matt}, {Matute}, {Miyaji}, {Molendi},
  {Paltani}, {Sanders}, {Scoville}, {Tresse}, {Urry}, {Vettolani}, \&
  {Zamorani}}]{2007ApJS..172...29H}
{Hasinger}, G., {Cappelluti}, N., {Brunner}, H., {et~al.} 2007, \apjs, 172, 29

\bibitem[{{Hatziminaoglou} {et~al.}(2008){Hatziminaoglou}, {Fritz},
  {Franceschini}, {Afonso-Luis}, {Hern{\'a}n-Caballero}, {P{\'e}rez-Fournon},
  {Serjeant}, {Lonsdale}, {Oliver}, {Rowan-Robinson}, {Shupe}, {Smith}, \&
  {Surace}}]{2008MNRAS.386.1252H}
{Hatziminaoglou}, E., {Fritz}, J., {Franceschini}, A., {et~al.} 2008, \mnras,
  386, 1252

\bibitem[{{Hatziminaoglou} {et~al.}(2009){Hatziminaoglou}, {Fritz}, \&
  {Jarrett}}]{2009MNRAS.tmp.1247H}
{Hatziminaoglou}, E., {Fritz}, J., \& {Jarrett}, T.~H. 2009, \mnras, 1247

\bibitem[{{H{\"o}nig} {et~al.}(2006){H{\"o}nig}, {Beckert}, {Ohnaka}, \&
  {Weigelt}}]{2006A&A...452..459H}
{H{\"o}nig}, S.~F., {Beckert}, T., {Ohnaka}, K., \& {Weigelt}, G. 2006, \aap,
  452, 459

\bibitem[{{Hopkins} {et~al.}(2006){Hopkins}, {Hernquist}, {Cox}, {Di Matteo},
  {Robertson}, \& {Springel}}]{2006ApJS..163....1H}
{Hopkins}, P.~F., {Hernquist}, L., {Cox}, T.~J., {et~al.} 2006, \apjs, 163, 1

\bibitem[{{Hopkins} {et~al.}(2007){Hopkins}, {Richards}, \&
  {Hernquist}}]{2007ApJ...654..731H}
{Hopkins}, P.~F., {Richards}, G.~T., \& {Hernquist}, L. 2007, \apj, 654, 731

\bibitem[{{Jaffe} {et~al.}(2004){Jaffe}, {Meisenheimer}, {R{\"o}ttgering},
  {Leinert}, {Richichi}, {Chesneau}, {Fraix-Burnet}, {Glazenborg-Kluttig},
  {Granato}, {Graser}, {Heijligers}, {K{\"o}hler}, {Malbet}, {Miley},
  {Paresce}, {Pel}, {Perrin}, {Przygodda}, {Schoeller}, {Sol}, {Waters},
  {Weigelt}, {Woillez}, \& {de Zeeuw}}]{2004Natur.429...47J}
{Jaffe}, W., {Meisenheimer}, K., {R{\"o}ttgering}, H.~J.~A., {et~al.} 2004,
  \nat, 429, 47

\bibitem[{{Kuraszkiewicz} {et~al.}(2009){Kuraszkiewicz}, {Wilkes}, {Schmidt},
  {Ghosh}, {Smith}, {Cutri}, {Hines}, {Huff}, {McDowell}, \&
  {Nelson}}]{2009ApJ...692.1143K}
{Kuraszkiewicz}, J., {Wilkes}, B.~J., {Schmidt}, G., {et~al.} 2009, \apj, 692,
  1143

\bibitem[{{Kuraszkiewicz} {et~al.}(2003){Kuraszkiewicz}, {Wilkes}, {Hooper},
  {McLeod}, {Wood}, {Bjorkman}, {Delain}, {Hughes}, {Elvis}, {Impey},
  {Lonsdale}, {Malkan}, {McDowell}, \& {Whitney}}]{2003ApJ...590..128K}
{Kuraszkiewicz}, J.~K., {Wilkes}, B.~J., {Hooper}, E.~J., {et~al.} 2003, \apj,
  590, 128

\bibitem[{{Lampton} {et~al.}(1976){Lampton}, {Margon}, \&
  {Bowyer}}]{1976ApJ...208..177L}
{Lampton}, M., {Margon}, B., \& {Bowyer}, S. 1976, \apj, 208, 177

\bibitem[{{Lusso} {et~al.}(2009){Lusso}, {Comastri}, {Vignali}, {Zamorani},
  {Brusa}, {Gilli}, {Iwasawa}, {Salvato}, {Civano}, {Elvis}, {Merloni},
  {Bongiorno}, {Trump}, {Koekemoer}, {Schinnerer}, {Le Floc'h}, {Cappelluti},
  {Jahnke}, {Sargent}, {Silverman}, {Mainieri}, {Fiore}, {Bolzonella}, {Le
  F{\`e}vre}, {Garilli}, {Iovino}, {Kneib}, {Lamareille}, {Lilly}, {Mignoli},
  {Scodeggio}, \& {Vergani}}]{2009arXiv0912.4166L}
{Lusso}, E., {Comastri}, A., {Vignali}, C., {et~al.} 2009, ArXiv e-prints

\bibitem[{{Maiolino} {et~al.}(2006){Maiolino}, {Mignoli}, {Pozzetti},
  {Severgnini}, {Brusa}, {Vignali}, {Puccetti}, {Ciliegi}, {Cocchia},
  {Comastri}, {Fiore}, {La Franca}, {Matt}, {Molendi}, \&
  {Perola}}]{2006A&A...445..457M}
{Maiolino}, R., {Mignoli}, M., {Pozzetti}, L., {et~al.} 2006, \aap, 445, 457

\bibitem[{{Maiolino} {et~al.}(2007){Maiolino}, {Shemmer}, {Imanishi}, {Netzer},
  {Oliva}, {Lutz}, \& {Sturm}}]{2007A&A...468..979M}
{Maiolino}, R., {Shemmer}, O., {Imanishi}, M., {et~al.} 2007, \aap, 468, 979

\bibitem[{{Makovoz} \& {Marleau}(2005)}]{2005PASP..117.1113M}
{Makovoz}, D. \& {Marleau}, F.~R. 2005, \pasp, 117, 1113

\bibitem[{{Marconi} \& {Hunt}(2003)}]{2003ApJ...589L..21M}
{Marconi}, A. \& {Hunt}, L.~K. 2003, \apjl, 589, L21

\bibitem[{{Marconi} {et~al.}(2004){Marconi}, {Risaliti}, {Gilli}, {Hunt},
  {Maiolino}, \& {Salvati}}]{2004MNRAS.351..169M}
{Marconi}, A., {Risaliti}, G., {Gilli}, R., {et~al.} 2004, \mnras, 351, 169

\bibitem[{{Mart{\'{\i}}nez-Sansigre} {et~al.}(2005){Mart{\'{\i}}nez-Sansigre},
  {Rawlings}, {Lacy}, {Fadda}, {Marleau}, {Simpson}, {Willott}, \&
  {Jarvis}}]{2005Natur.436..666M}
{Mart{\'{\i}}nez-Sansigre}, A., {Rawlings}, S., {Lacy}, M., {et~al.} 2005,
  \nat, 436, 666

\bibitem[{{McLure} \& {Dunlop}(2004)}]{2004MNRAS.352.1390M}
{McLure}, R.~J. \& {Dunlop}, J.~S. 2004, \mnras, 352, 1390

\bibitem[{{Merloni} {et~al.}(2010){Merloni}, {Bongiorno}, {Bolzonella},
  {Brusa}, {Civano}, {Comastri}, {Elvis}, {Fiore}, {Gilli}, {Hao}, {Jahnke},
  {Koekemoer}, {Lusso}, {Mainieri}, {Mignoli}, {Miyaji}, {Renzini}, {Salvato},
  {Silverman}, {Trump}, {Vignali}, {Zamorani}, {Capak}, {Lilly}, {Sanders},
  {Taniguchi}, {Bardelli}, {Carollo}, {Caputi}, {Contini}, {Coppa}, {Cucciati},
  {de la Torre}, {de Ravel}, {Franzetti}, {Garilli}, {Hasinger}, {Impey},
  {Iovino}, {Iwasawa}, {Kampczyk}, {Kneib}, {Knobel}, {Kova{\v c}},
  {Lamareille}, {Le Borgne}, {Le Brun}, {Le F{\`e}vre}, {Maier}, {Pello},
  {Peng}, {Perez Montero}, {Ricciardelli}, {Scodeggio}, {Tanaka}, {Tasca},
  {Tresse}, {Vergani}, \& {Zucca}}]{2010ApJ...708..137M}
{Merloni}, A., {Bongiorno}, A., {Bolzonella}, M., {et~al.} 2010, \apj, 708, 137

\bibitem[{{Mignoli} {et~al.}(2004){Mignoli}, {Pozzetti}, {Comastri}, {Brusa},
  {Ciliegi}, {Cocchia}, {Fiore}, {La Franca}, {Maiolino}, {Matt}, {Molendi},
  {Perola}, {Puccetti}, {Severgnini}, \& {Vignali}}]{2004A&A...418..827M}
{Mignoli}, M., {Pozzetti}, L., {Comastri}, A., {et~al.} 2004, \aap, 418, 827

\bibitem[{{Nenkova} {et~al.}(2002){Nenkova}, {Ivezi{\'c}}, \&
  {Elitzur}}]{2002ApJ...570L...9N}
{Nenkova}, M., {Ivezi{\'c}}, {\v Z}., \& {Elitzur}, M. 2002, \apjl, 570, L9

\bibitem[{{Nenkova} {et~al.}(2008){Nenkova}, {Sirocky}, {Ivezi{\'c}}, \&
  {Elitzur}}]{2008ApJ...685..147N}
{Nenkova}, M., {Sirocky}, M.~M., {Ivezi{\'c}}, {\v Z}., \& {Elitzur}, M. 2008,
  \apj, 685, 147

\bibitem[{{Perola} {et~al.}(2004){Perola}, {Puccetti}, {Fiore}, {Sacchi},
  {Brusa}, {Cocchia}, {Baldi}, {Carangelo}, {Ciliegi}, {Comastri}, {La Franca},
  {Maiolino}, {Matt}, {Mignoli}, {Molendi}, \& {Vignali}}]{2004A&A...421..491P}
{Perola}, G.~C., {Puccetti}, S., {Fiore}, F., {et~al.} 2004, \aap, 421, 491

\bibitem[{{Peterson} {et~al.}(2004){Peterson}, {Ferrarese}, {Gilbert}, {Kaspi},
  {Malkan}, {Maoz}, {Merritt}, {Netzer}, {Onken}, {Pogge}, {Vestergaard}, \&
  {Wandel}}]{2004ApJ...613..682P}
{Peterson}, B.~M., {Ferrarese}, L., {Gilbert}, K.~M., {et~al.} 2004, \apj, 613,
  682

\bibitem[{{Pier} \& {Krolik}(1992)}]{1992ApJ...401...99P}
{Pier}, E.~A. \& {Krolik}, J.~H. 1992, \apj, 401, 99

\bibitem[{{Pier} \& {Krolik}(1993)}]{1993ApJ...418..673P}
{Pier}, E.~A. \& {Krolik}, J.~H. 1993, \apj, 418, 673

\bibitem[{{Polletta} {et~al.}(2008){Polletta}, {Weedman}, {H{\"o}nig},
  {Lonsdale}, {Smith}, \& {Houck}}]{2008ApJ...675..960P}
{Polletta}, M., {Weedman}, D., {H{\"o}nig}, S., {et~al.} 2008, \apj, 675, 960

\bibitem[{{Pozzi} {et~al.}(2007){Pozzi}, {Vignali}, {Comastri}, {Pozzetti},
  {Mignoli}, {Gruppioni}, {Zamorani}, {Lari}, {Civano}, {Brusa}, {Fiore},
  {Maiolino}, \& {La Franca}}]{2007A&A...468..603P}
{Pozzi}, F., {Vignali}, C., {Comastri}, A., {et~al.} 2007, \aap, 468, 603

\bibitem[{{Renzini}(2006)}]{2006ARA&A..44..141R}
{Renzini}, A. 2006, \araa, 44, 141

\bibitem[{{Richards} {et~al.}(2006){Richards}, {Lacy}, {Storrie-Lombardi},
  {Hall}, {Gallagher}, {Fan}, {Papovich}, {Vanden Berk}, {Trammell},
  {Schneider}, {Vestergaard}, {York}, {Jester}, {Anderson}, {Budav{\'a}ri}, \&
  {Szalay}}]{2006ApJS..166..470R}
{Richards}, G.~T., {Lacy}, M., {Storrie-Lombardi}, L.~J., {et~al.} 2006, \apjs,
  166, 470

\bibitem[{{Rigby} {et~al.}(2005){Rigby}, {Rieke}, {P{\'e}rez-Gonz{\'a}lez},
  {Donley}, {Alonso-Herrero}, {Huang}, {Barmby}, \&
  {Fazio}}]{2005ApJ...627..134R}
{Rigby}, J.~R., {Rieke}, G.~H., {P{\'e}rez-Gonz{\'a}lez}, P.~G., {et~al.} 2005,
  \apj, 627, 134

\bibitem[{{Risaliti} {et~al.}(2002){Risaliti}, {Elvis}, \&
  {Nicastro}}]{2002ApJ...571..234R}
{Risaliti}, G., {Elvis}, M., \& {Nicastro}, F. 2002, \apj, 571, 234

\bibitem[{{Rodighiero} {et~al.}(2007){Rodighiero}, {Gruppioni}, {Civano},
  {Comastri}, {Franceschini}, {Mignoli}, {Fritz}, {Vignali}, \&
  {Treu}}]{2007MNRAS.376..416R}
{Rodighiero}, G., {Gruppioni}, C., {Civano}, F., {et~al.} 2007, \mnras, 376,
  416

\bibitem[{{Salpeter}(1955)}]{1955VA......1..283S}
{Salpeter}, E.~E. 1955, Vistas in Astronomy, 1, 283

\bibitem[{{Shen} {et~al.}(2008){Shen}, {Greene}, {Strauss}, {Richards}, \&
  {Schneider}}]{2008ApJ...680..169S}
{Shen}, Y., {Greene}, J.~E., {Strauss}, M.~A., {Richards}, G.~T., \&
  {Schneider}, D.~P. 2008, \apj, 680, 169

\bibitem[{{Shields} {et~al.}(2006){Shields}, {Salviander}, \&
  {Bonning}}]{2006NewAR..50..809S}
{Shields}, G.~A., {Salviander}, S., \& {Bonning}, E.~W. 2006, New Astronomy
  Review, 50, 809

\bibitem[{{Silva} {et~al.}(2004){Silva}, {Maiolino}, \&
  {Granato}}]{2004MNRAS.355..973S}
{Silva}, L., {Maiolino}, R., \& {Granato}, G.~L. 2004, \mnras, 355, 973

\bibitem[{{Spergel} {et~al.}(2003){Spergel}, {Verde}, {Peiris}, {Komatsu},
  {Nolta}, {Bennett}, {Halpern}, {Hinshaw}, {Jarosik}, {Kogut}, {Limon},
  {Meyer}, {Page}, {Tucker}, {Weiland}, {Wollack}, \&
  {Wright}}]{2003ApJS..148..175S}
{Spergel}, D.~N., {Verde}, L., {Peiris}, H.~V., {et~al.} 2003, \apjs, 148, 175

\bibitem[{{Stark} {et~al.}(1992){Stark}, {Gammie}, {Wilson}, {Bally}, {Linke},
  {Heiles}, \& {Hurwitz}}]{1992ApJS...79...77S}
{Stark}, A.~A., {Gammie}, C.~F., {Wilson}, R.~W., {et~al.} 1992, \apjs, 79, 77

\bibitem[{{Steffen} {et~al.}(2006){Steffen}, {Strateva}, {Brandt}, {Alexander},
  {Koekemoer}, {Lehmer}, {Schneider}, \& {Vignali}}]{2006AJ....131.2826S}
{Steffen}, A.~T., {Strateva}, I., {Brandt}, W.~N., {et~al.} 2006, \aj, 131,
  2826

\bibitem[{{Vasudevan} \& {Fabian}(2009)}]{2009MNRAS.392.1124V}
{Vasudevan}, R.~V. \& {Fabian}, A.~C. 2009, \mnras, 392, 1124

\bibitem[{{Vasudevan} {et~al.}(2009){Vasudevan}, {Fabian}, {Gandhi}, {Winter},
  \& {Mushotzky}}]{2009arXiv0910.5256V}
{Vasudevan}, R.~V., {Fabian}, A.~C., {Gandhi}, P., {Winter}, L.~M., \&
  {Mushotzky}, R.~F. 2009, ArXiv e-prints

\bibitem[{{Vignali} {et~al.}(2003){Vignali}, {Brandt}, \&
  {Schneider}}]{2003AJ....125..433V}
{Vignali}, C., {Brandt}, W.~N., \& {Schneider}, D.~P. 2003, \aj, 125, 433

\bibitem[{{Vignali} {et~al.}(2009){Vignali}, {Pozzi}, {Fritz}, {Comastri},
  {Gruppioni}, {Bellocchi}, {Fiore}, {Brusa}, {Maiolino}, {Mignoli}, {La
  Franca}, {Pozzetti}, {Zamorani}, \& {Merloni}}]{2009MNRAS.395.2189V}
{Vignali}, C., {Pozzi}, F., {Fritz}, J., {et~al.} 2009, \mnras, 395, 2189 (V09)

\bibitem[{{Weedman} {et~al.}(2006){Weedman}, {Polletta}, {Lonsdale}, {Wilkes},
  {Siana}, {Houck}, {Surace}, {Shupe}, {Farrah}, \&
  {Smith}}]{2006ApJ...653..101W}
{Weedman}, D., {Polletta}, M., {Lonsdale}, C.~J., {et~al.} 2006, \apj, 653, 101

\bibitem[{{Witt} {et~al.}(1997){Witt}, {Czerny}, \&
  {Zycki}}]{1997MNRAS.286..848W}
{Witt}, H.~J., {Czerny}, B., \& {Zycki}, P.~T. 1997, \mnras, 286, 848

\bibitem[{{Zamorani} {et~al.}(1981){Zamorani}, {Henry}, {Maccacaro},
  {Tananbaum}, {Soltan}, {Avni}, {Liebert}, {Stocke}, {Strittmatter},
  {Weymann}, {Smith}, \& {Condon}}]{1981ApJ...245..357Z}
{Zamorani}, G., {Henry}, J.~P., {Maccacaro}, T., {et~al.} 1981, \apj, 245, 357

\bibitem[{{Zombeck}(1990)}]{1990hsaa.book.....Z}
{Zombeck}, M.~V. 1990, {Handbook of space astronomy and astrophysics}, ed.
  M.~V. Zombeck

\end{thebibliography}

\end{document}